\documentclass[twocolumn, prb, aps, superscriptaddress, longbibliography, floatfix]{revtex4-1}
\pdfoutput=1
\usepackage[caption=false,subrefformat=parens,labelformat=parens]{subfig}
\usepackage{graphicx}
\usepackage{multirow}
\usepackage{float}
\usepackage{amsmath}
\usepackage{bm}
\usepackage{xcolor}
\usepackage{siunitx}
\usepackage{tabularx}
\usepackage[colorlinks=true, allcolors=blue]{hyperref}
\usepackage{color}
\usepackage{dsfont}

\begin{document}

\title{Superconducting gap symmetry from Bogoliubov quasiparticle interference analysis on \texorpdfstring{{Sr}$_2${RuO}$_4$}{Lg}}

\author{Shinibali Bhattacharyya}
\affiliation{Institut f\"ur Theoretische Physik, Goethe-Universit\"at, 60438 Frankfurt am Main, Germany}
\author{Andreas Kreisel}
\affiliation{Institut f\" ur Theoretische Physik, Universit\"at Leipzig, D-04103 Leipzig, Germany}
\author{X. Kong}
\affiliation{Center For Nanophase Materials Sciences, Oak Ridge National Laboratory, Oak Ridge, Tennessee 37831, USA}
\author{T. Berlijn}
\affiliation{Center For Nanophase Materials Sciences, Oak Ridge National Laboratory, Oak Ridge, TN 37831, USA}
\author{Astrid T. R{\o}mer}
\affiliation{
Niels Bohr Institute, University of Copenhagen, R\aa dmandsgade 62, DK-2200 Copenhagen, Denmark}
\author{Brian M. Andersen}
\affiliation{
Niels Bohr Institute, University of Copenhagen, R\aa dmandsgade 62, DK-2200 Copenhagen, Denmark}
\author{P. J. Hirschfeld}
\affiliation{Department of Physics, University of Florida, Gainesville, Florida 32611, USA}

\date{\today}
\begin{abstract}

The nature of the superconducting order parameter
in {Sr}$_2${RuO}$_4$ has generated intense  interest in recent years. Since  the superconducting gap  is very small, high resolution methods such as scanning tunneling spectroscopy might be the best chance to directly resolve the gap symmetry. Recently, a Bogoliubov quasiparticle interference imaging (BQPI) experiment has suggested that the $d_{x^2-y^2}$ gap symmetry is appropriate for {Sr}$_2${RuO}$_4$. In this work, we use a material-specific theoretical approach based on Wannier functions of the surface of {Sr}$_2${RuO}$_4$ to calculate the continuum density of states as detected in scanning tunneling microscopy experiments. We examine several different proposed gap order parameters, and calculate the expected BQPI pattern for each case. Comparing to the available experimental data, our results suggest that a $s'+id_{xy}$ gap order parameter is the most probable state, but the measured BQPI patterns still display features unaccounted for by the theory for any of the states currently under discussion.
\end{abstract}


\maketitle
\section{Introduction}

{Sr}$_2${RuO}$_4$ \citep{Mackenzie_RMP_2003,Maeno_JPSJ_2012,Kallin_IOP_2016,Mackenzie_npj_2017} has once again been a topic of intense research interest recently.
For long time it was discussed as a promising candidate for intrinsic topological superconductor in the light of arguments in favor of  a $p+ip$ symmetry of the gap order, in particular early evidence from NMR\citep{Ishida_Nature_1998}.
Muon-spin rotation \citep{Luke_Nature_1998} and Kerr effect \citep{Xia_PRL_2006} measurements were argued to imply time-reversal symmetry (TRS) breaking \citep{Rice_1995} in the superconducting state, consistent with this picture.
However, a number of thermodynamic measurements detected  low-energy quasiparticle states,
hinting at the existence of nodes or deep minima in the gap $\Delta(\mathbf{k})$\cite{Bonalde_PRL_2000,Mao_JPSJ_2000,Maeno_PRL_2004,Hassinger_PRX_2017}.

Recently, strong evidence against the chiral $p$-wave paradigm was provided by in-plane $^{17}$O nuclear magnetic resonance measurements \citep{Pustogow_Nature_2019,Ishida_correct} that found a substantial drop in the {K}night shift below $T_\mathrm{c}$.  Arguments in favor of spin-singlet pairing were then drawn from the field dependence of the {K}night shift in comparison to the change of the entropy from earlier specific heat experiments\citep{Brown_2020}. These measurements are complemented by observations of shifts in the elastic constants\cite{c66_ref,ghosh2020thermodynamic} together with experiments under strain\cite{Grinenko2021} that indicate a two component nature of the order parameter.
All these findings have led to a series of recent theoretical attacks on the question of  superconductivity in {Sr}$_2${RuO}$_4$ \citep{Astrid_PRL_2019,Scaffidi_PRR_2019,Gingras_PRL_2019,suh2019,Kaba19,Ramires2019,Acharya_CommPhys_2019,Kallin_PRB_2020, Romer2020, romer2020fluctuationdriven,Romer2021,kivelson_npj,clepkens2021,willa2021,Astrid3D2022,Merce2022,Henrik2022}.

{Sr}$_2${RuO}$_4$ is generally considered to be an unconventional superconductor, where electron pairing is mediated by the exchange of electronic excitations, so theories typically attempt to model the low-energy effective pairing interaction. The  Fermi surface is dominated by
three Ru $d$ orbitals $d_{xz},d_{yz}$ that contribute mostly to quasi-1D bands, and $d_{xy}$ states dominating a 2D band. Electrons in these states interact via  intrasite Coulomb $U$ interactions and Hund's coupling $J$, with the $d_{xy}$ dominant bands thought to be more strongly correlated.~\cite{Kugler2020,Mravlje11} In addition, spin-orbit coupling plays a significant role\citep{Damascelli_PRL_2008,Georges_PRL_2018}. {Recent microscopic theories incorporate many of these ingredients \citep{Astrid_PRL_2019,Romer2021,Scaffidi_PRR_2019,Gingras_PRL_2019,Acharya_CommPhys_2019,Kallin_PRB_2020,Raghu_PRL_2010,Scaffidi_PRB_2014,Astrid3D2022,kivelson_npj,Ramires2019,Kaba19,Astrid_PRL_2019,Scaffidi_PRR_2019,Gingras_PRL_2019,willa2021}, leading to a variety of predictions for $\Delta(\mathbf{k})$ depending on the assumed model and the applied methodology. These calculations point to a number of leading candidates which currently consist of the even-parity 1D irreducible representation $B_{1g} (d_{x^2-y^2})$, multi-component orders such as 
$d_{x^2-y^2}+ig_{xy(x^2-y^2)}$ and $s'+id_{xy}$, as well as the 2D irreducible representation $E_{1g} (d_{xz}+id_{yz})$. A two-component state is generally thought to be important to explain the observation of time-reversal symmetry breaking, ultrasound\cite{Benhabib2021,ghosh2020thermodynamic}, and recent $\mu$SR experiments under strain\cite{Grinenko2021}.  At zero strain, these can  correspond either to the 2D representation, or to accidental degeneracies of two 1D representations.}

Not all such proposals are consistent with existing experimental results, nor are all the interpretations of the experimental literature apparently consistent with one another\cite{Mackenzie_npj_2017}.  In such a situation it would be very useful to have a direct measurement of the superconducting gap to distinguish among theories and thereby constrain  the possible pairing mechanisms. 
The tiny size ($|\Delta|\leq350 \, \mu$eV)~\citep{Firmo2013,Madhavan_PNAS_2020} of the superconducting order parameter $\Delta(\mathbf{k})$ in {Sr}$_2${RuO}$_4$ has hindered such measurements for a long time, as they require low temperatures and fine energy resolution to detect spectral features arising from the small gap.  However, in principle Bogoliubov quasiparticle interference (BQPI) imaging is a powerful technique capable of high-precision measurement of multiband $\Delta(\mathbf{k})$ \citep{Nunner2006,Hanaguri_NatPhys_2007,Allan_Science_2012,Sprau2017}.
Interference of impurity-scattered quasiparticles produces real space Friedel oscillations in the density of states (DOS) $\rho(\mathbf{r},\omega)$ within the energy range of $\Delta(\mathbf{k})$, giving rise to complex patterns in the spatial Fourier transform spectrum $\rho({\bf q},\omega)$ that can in principle   be measured in an  STM experiment and interpreted in order to extract the gap structure.
The BQPI technique was recently implemented to analyze $\Delta(\mathbf{k})$ in {Sr}$_2${RuO}$_4$ \citep{Madhavan_PNAS_2020}. This analysis, motivated with observations from recent work \citep{Hassinger_PRX_2017,Pustogow_Nature_2019,Maeno_PRB_2019,Astrid_PRL_2019,Scaffidi_PRR_2019,Gingras_PRL_2019}, suggested a $d_{x^2-y^2}$ superconducting gap symmetry for {Sr}$_2${RuO}$_4$ by comparison with theoretical calculations of $\rho({\bf q},\omega)$ using simple low-harmonic gap candidates $\Delta({\bf k})$.  Such an approach is, however, limited by a) loss of ${\bf q}$ space resolution  when calculating BQPI on a simple Ru lattice system and b) failure to include the  rather complex gap structures, including accidental nodes, anticipated by microscopic theories\cite{Astrid_PRL_2019,Romer2021,Astrid3D2022}.

In this work, we will adopt the  Wannier-based $T$-matrix technique introduced in Refs.~\onlinecite{Kreisel_PRL_2015,Kreisel_PRB_2016} through which one obtains BQPI images for real materials in its normal and superconducting states, that are directly comparable to experiments. Unlike the theoretical approach used in Ref.~\onlinecite{Madhavan_PNAS_2020} to obtain lattice DOS $\rho(\mathbf{R},\omega)$, the Wannier-based $T$-matrix approach is used for evaluation of the continuum DOS $\rho(\mathbf{r},\omega)$ dressed by real-space Wannier functions pertaining to the orbitals that have dominant contribution around the Fermi level of the material under consideration. These Wannier functions substantially modify the continuum DOS patterns, bringing in additional $\textbf{q}(\omega)$ features in its Fourier transformed image.  In addition, we take as representative of the various possible gap symmetries not only simple harmonics, but gaps derived from microscopic spin fluctuation pairing theory\cite{Romer2021}. We find that definitive conclusions are difficult because of limited experimental    tunneling spectra  exhibiting superconductivity on the surface (i.e. with evidence of low-energy coherence peaks), and the  fact that BQPI features are remarkably nondispersive in this system, even for highly anisotropic gap structures. The best fit to the data currently available\citep{Madhavan_PNAS_2020} suggests a $s'+id_{xy}$ gap order parameter, but we discuss other alternatives in detail.

This paper is organized as follows: in Sec \ref{Sec:model}, we introduce the multiorbital Hubbard Hamiltonian with the pairing and impurity terms, and describe it using the Bardeen-Cooper-Schrieffer (BCS) equations in the Nambu spinor basis. We write down the equations for computing the lattice DOS and continuum DOS \citep{Kreisel_PRL_2015,Kreisel_PRB_2016}, in the presence of impurities using the $T$-matrix approach. In Sec \ref{Sec:results}, we demonstrate our BQPI findings first in the normal state, and second in the superconducting state of the material for which we borrow the singlet and composite gap order parameters as obtained in Refs.~\onlinecite{Astrid_PRL_2019, Romer2021}. Finally, we present our conclusions in Sec \ref{Sec:summary} and possible future directions of investigation in this context.

\section{Model} \label{Sec:model}

\begin{figure*}[bt]
    \centering
    \includegraphics[width=\textwidth]{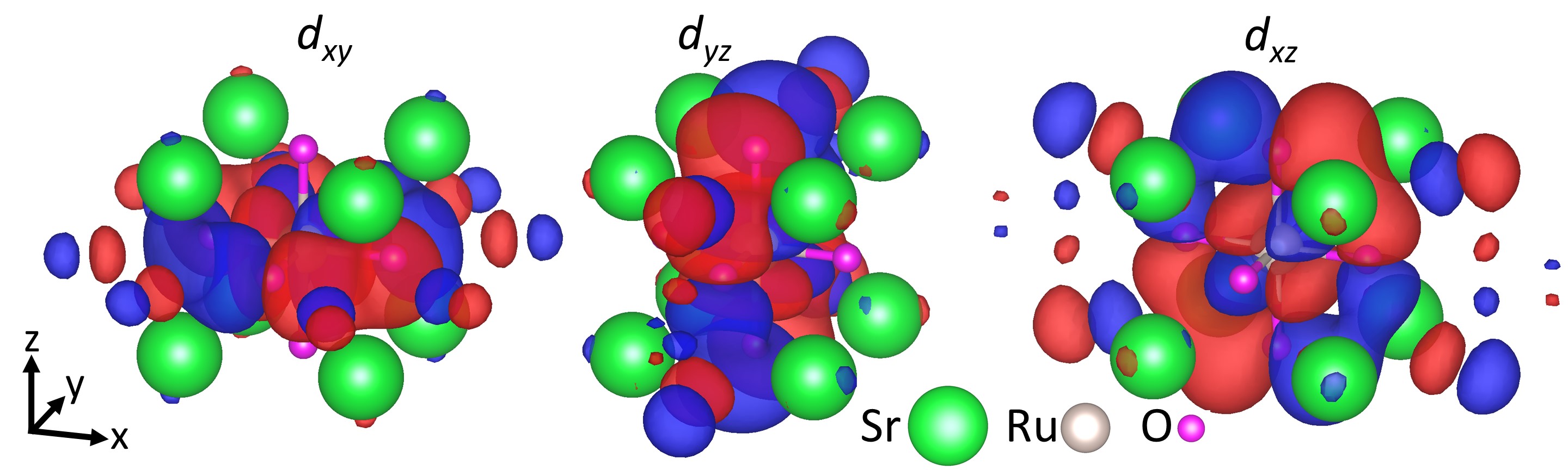}
	\caption{Isosurface plots of Ru $d_{xy}$, $d_{yz}$ and $d_{xz}$ Wannier orbitals in {Sr}$_2${RuO}$_4$. The atoms Sr, Ru and O are depicted by the opaque spheres colored in green, white and pink, respectively. The Ru atom is positioned at the center of the unit cell which has an in-plane lattice constant of $a = 3.8644$ \AA{} as considered in the {\it ab-initio} calculations described in Appendix \ref{Appendix:wannier}. The Wannier functions are depicted by the translucent red and blue colored lobes indicating opposite phases of the wave functions.}
    \label{Fig:SROwannier}
\end{figure*}
In this section, we will lay out the theoretical framework on how the  surface of a material is imaged by STM.
The full Hamiltonian has four terms, namely the kinetic energy term $H_{0}$, the spin-orbit term $H_{\textnormal{soc}}$, the BCS mean-field term $H_{\textnormal{MF}}$ and a single-impurity term $H_{\textnormal{imp}}$.
\begin{eqnarray}
\label{Eqn:FullH_with_Himp}
H &=& H_{0} + H_{\textnormal{soc}} + H_{\textnormal{MF}} + H_{\textnormal{imp} }, \\
H_{0} &=& \sum_{\mathbf{R} \mathbf{R^\prime} m n s} t_{\mathbf{R} \mathbf{R^\prime}}^{m n} c_{\mathbf{R} m  s}^{\dagger} c_{\mathbf{R^\prime} n s} - \mu _{0}\sum_{\mathbf{R} m s}c_{\mathbf{R} m  s}^{\dagger} c_{\mathbf{R} m s}, \nonumber \\
H_{\textnormal{MF}} &=& - \sum_{\mathbf{R} \mathbf{R^\prime} m n } [\Delta _{\mathbf{R} \mathbf{R^\prime}}^{m n}c_{\mathbf{R} m  \uparrow}^{\dagger}c_{\mathbf{R^\prime} n  \downarrow}^{\dagger} + h.c.],  \nonumber \\
H_{\textnormal{imp}} &=&  \sum_{\mathbf{R}^{\star} m n s} V_{\textnormal{imp}}^{m n} c_{\mathbf{R}^{\star} m  s}^{\dagger} c_{\mathbf{R}^{\star} n s}.   \nonumber
\end{eqnarray}
$H_{\textnormal{soc}}$ will be introduced in the next subsection. Here, $c_{\mathbf{R} m  s}^{\dagger}$ ($c_{\mathbf{R} m  s}$) is the creation (annihilation) operator for an electron in the unit cell $\mathbf{R}$, orbital $m$ with spin $s$. $t_{\mathbf{R} \mathbf{R^\prime}}^{m n}$ is the amplitude for hopping from unit cell $\mathbf{R}$, orbital $m$ to the unit cell $\mathbf{R^\prime}$, orbital $n$. The pairing field is given by $\Delta _{\mathbf{R} \mathbf{R^\prime}}^{m n} = V_{\mathbf{R} \mathbf{R^\prime}}^{m n} \left\langle c_{\mathbf{R^\prime} n  \downarrow} c_{\mathbf{R} m  \uparrow}  \right\rangle $, where
$V _{\mathbf{R} \mathbf{R^\prime}}^{m n}$ is the effective attraction between unit cell $\mathbf{R}$, orbital $m$ and unit cell $\mathbf{R^\prime}$, orbital $n$. $\mathbf{R}^{\star}$ is the impurity site and $ V_{\textnormal{imp}}^{m n}$ is the (on-site only) non-magnetic impurity potential responsible for scattering of quasiparticles.

\subsection{Non-interacting Hamiltonian}
\label{Subsec:non-int-H}
We will adopt the non-interacting tight-binding Hamiltonian model for {Sr}$_2${RuO}$_4$ as used in Refs.~\onlinecite{Astrid_PRL_2019,Romer2020,Romer2021}. It is composed of the three Ru orbitals: $d_{xz}$, $d_{yz}$ and $d_{xy}$. Fourier transforming the real space hoppings from $H_0$ fitted to ARPES measurements \citep{Zabolotny2013,Cobo_2016}, one obtains the dispersions given by $\xi_{xz}(\mathbf{k}) = -2t_1 \cos \, k_x -2t_2 \cos \, k_y - \mu$, $\xi_{yz}(\mathbf{k}) = -2t_2 \cos \, k_x -2t_1 \cos \, k_y - \mu$, and $\xi_{xy}(\mathbf{k}) = -2t_3( \cos \, k_x + \cos \, k_y) -4t_4 \cos \, k_x \cos \, k_y -2t_5( \cos \, 2k_x + \cos \, 2k_y) - \mu$ with $\{t_1,t_2,t_3,t_4,t_5\} = \{ 88,9,80,40,5,109\}$ meV with the inter-orbital hybridization $g(\textbf{k})$ set to 0 (see Eq.~(\ref{Eqn:H0}) below) and the chemical potential $\mu = 109$ meV. Atomic SOC is parametrized by $H_{\textnormal{soc}} = \lambda_{\textnormal{soc}} \textbf{L}\cdot\textbf{S}$ and gives rise to orbital mixing on the Fermi surface sheets. We include SOC of $\lambda_{\textnormal{soc}} \approx 40$ meV $(\approx 0.5t_1)$ \citep{Astrid_PRL_2019, Romer2020, Romer2021}. For the TRS-preserved normal state, we have doubly degenerate Kramer's eigenvalues and the non-interacting Hamiltonian $\hat{H}$ in block-diagonal form can be represented in a pseudospin basis $\sigma = + (-)$. Here $\hat{H} = \sum_{\textbf{k}\sigma} \Psi^\dagger(\textbf{k}, \sigma) (H_0 + H_{\textnormal{soc}}(\sigma)) \Psi(\textbf{k}, \sigma)$ where,
\begin{align}
    \label{Eqn:H0}
    H_0 = &\left( \begin{array}{ccc} 
    \xi_{xz}(\textbf{k}) & g(\textbf{k}) & 0 \\
    g(\textbf{k}) & \xi_{yz}(\textbf{k}) & 0 \\
     0 & 0 & \xi_{xy}(\textbf{k}) \\
    \end{array} 
    \right), \\
    \label{Eqn:HSOC}
    H_{\textnormal{soc}}(\sigma) = \frac{1}{2} & \left( \begin{array}{ccc} 
    0 & -i\sigma\lambda_{\textnormal{soc}} & i\lambda_{\textnormal{soc}} \\
    i\sigma\lambda_{\textnormal{soc}} & 0 & -\sigma\lambda_{\textnormal{soc}} \\
    -i\lambda_{\textnormal{soc}} & -\sigma\lambda_{\textnormal{soc}} & 0 \\
    \end{array} 
    \right),
\end{align}
and $\Psi(\textbf{k}, +) = [c_{xz\uparrow}(\textbf{k}), c_{yz\uparrow}(\textbf{k}), c_{xy\downarrow}(\textbf{k})]^T$ and $\Psi(\textbf{k}, -) = [c_{xz\downarrow}(\textbf{k}), c_{yz\downarrow}(\textbf{k}), c_{xy\uparrow}(\textbf{k})]^T$.

\subsection{Superconducting gap} The full BCS Hamiltonian (implied by the underscore) is given by:
\begin{equation}
    \underline{\hat{H}} = \left( \begin{array}{cc} 
    \hat{H}(\mathbf{k}) & \hat{\Delta}(\mathbf{k}) \\
    \hat{\Delta}^\dagger(\mathbf{k}) & -\hat{H}^T(-\mathbf{k})
    \end{array} 
    \right),
    \label{Eqn:fullBCS}
\end{equation}
written in the Nambu spinor basis $ (\Phi^\dagger(\mathbf{k}),\Phi^T(-\mathbf{k}))$ where $\Phi^\dagger(\mathbf{k}) = (\Psi^\dagger(\textbf{k},+), \Psi^\dagger(\textbf{k},-)) $. The superconducting gap in momentum space in the homogeneous system $\hat{\Delta}(\mathbf{k})$ is obtained from the spin-fluctuation pairing theory evaluations\citep{Astrid_PRL_2019,Romer2021}. Diagonalizing $\underline{\hat{H}}$ yields the eigenvalues $\{ \pm E_\mu(\mathbf{k})\}$ and the unitary transformation matrix $\underline{\hat{U}}(\mathbf{k})$ that diagonalizes $\underline{\hat{H}}$.  The orbitally resolved-gap structure $\hat{\Delta}^{mn}(\mathbf{k})$ is also expressed in this Nambu basis\citep{Kreisel_PRB_2016} [see Appendix \ref{Appendix:Spatialgap} for theoretical details]. We evaluate six different cases of  pseudospin-singlet $(|\uparrow \downarrow \rangle - |\downarrow \uparrow \rangle)$ gap order parameters: simple intra-band $d^S_{x^2-y^2}$ where $\hat{\Delta}_\mu(\textbf{k}) = \frac{\Delta_0}{2}(\cos \, k_x - \cos \, k_y)$, and the rest taken from Refs.~\onlinecite{Astrid_PRL_2019, Romer2021}  $d_{x^2-y^2} $, $s' $, $s'+id_{x^2-y^2} $ and $s'+id_{xy} $. The spin-fluctuation mediated pairing theory for weakly-coupled systems (like Sr$_2$RuO$_4$) enables one to find the phase diagram of the possible leading gap order parameters as a function of the on-site Coulomb repulsion $U$ and Hund´s coupling $J$ interaction terms [\onlinecite{Astrid_PRL_2019}]. The gap solutions $d_{x^2-y^2} $, $s' $ and $s'+id_{x^2-y^2} $ are the results of such an analysis pertaining to different $U$ and $J$ values. Furthermore, Ref.~\onlinecite{Romer2021} also included the effect of nearest-neighbor Coulomb repulsion $V$ to investigate the fate of the leading gap order which yields a robust $s'+id_{xy} $ solution.

The orbitally-resolved homogeneous DOS in the superconducting state is $\rho^n(\omega)=-\frac{1}{\pi} \textnormal{Im} \sum_\mathbf{k} \hat{G}_{nn}(\mathbf{k},\omega)$ where
\begin{equation}
\underline{\hat{G}}(\mathbf{k},\omega) = (\omega - \underline{\hat{H}}(\mathbf{k}) + i\delta )^{-1},
\end{equation}
is the Green's function for the BCS Hamiltonian. The real-space bare lattice Green's function is obtained from the Fourier transform
\begin{equation}
\underline{\hat{G}}^0(\mathbf{R},\mathbf{R^\prime},\omega) = \sum_\mathbf{k} e^{-i \mathbf{k.}(\mathbf{R-R^\prime})} \underline{\hat{G}}^0(\mathbf{k},\omega) = \underline{\hat{G}}^0(\mathbf{R-R^\prime},\omega).
\end{equation}

\subsection{Impurity states and T-matrix approach} 
With intra-orbital, on-site impurity potential terms $V_{\textnormal{imp}}^{nn}$ in $H_{\textnormal{imp}}$ of Eq.~(\ref{Eqn:FullH_with_Himp}), one constructs the lattice Green's function in the presence of the impurity via the T-matrix approach:
\begin{align}
\underline{\hat{G}}(\mathbf{R},\mathbf{R^\prime},\omega) = & \underline{\hat{G}}^0(\mathbf{R}-\mathbf{R^\prime},\omega) + \underline{\hat{G}}^0(\mathbf{R},\omega) \underline{\hat{T}}(\omega) \underline{\hat{G}}^0(-\mathbf{R^\prime},\omega).
\label{Eqn:ImpurityGreen_Tmatrix}
\end{align}
The T-matrix is given by 
\begin{equation}
\underline{\hat{T}}(\omega) = [\mathds{1}-\underline{\hat{V}}_{\textnormal{imp}} \underline{\hat{G}} (\omega)]^{-1} \underline{\hat{V}}_{\textnormal{imp}},
\label{Eqn:Tmatrix}
\end{equation}
where the local Green's function is $\underline{\hat{G}}(\omega) = \sum_\mathbf{k} \underline{\hat{G}}(\mathbf{k},\omega)$ and the diagonal matrix $ \underline{\hat{V}}_{\textnormal{imp}} = V_{\textnormal{imp}}^{mn} \delta_{mn}\underline{\tau_z}$, implying its expression in the Nambu spinor basis with 
\begin{equation}
    \underline{\tau_z} = \left( \begin{array}{cc} 
    \mathds{1} & 0 \\
    0 & -\mathds{1}
    \end{array} 
    \right).
\end{equation}

\subsection{Wannier functions to calculate continuum density of states}

For a given bias voltage $V$, the differential tunneling conductance in an STM experiment is given by\citep{Hoffman2011}:
\begin{equation}
\frac{dI}{dV}(\mathbf{r},eV) = \frac{4 \pi e}{\hbar}\vert M \vert ^{2} \rho_{\text{tip}}(0) \rho(\mathbf{r},eV),
\label{Eqn:STMconductance}
\end{equation}
where $\mathbf{r} = (x,y,z)$ denote the coordinates of the tip, $\rho(\mathbf{r},eV)$ is
the continuum LDOS (cLDOS), $\rho_{\text{tip}}(0)$ is the DOS of the tip,
and $|M|^2$ is the square of the matrix element for the tunneling
barrier. The following methodology to include the effect of Wannier functions in modifying the cLDOS was introduced in Refs. \onlinecite{Kreisel_PRL_2015} and \onlinecite{Kreisel_PRB_2016}. The cLDOS can be calculated by
\begin{equation}
    \rho(\mathbf{r},\omega )= -\frac{1}{\pi}\textnormal{Im} G^{11}(\mathbf{r},\mathbf{r},\omega),
    \label{Eqn:cLDOS}
\end{equation}
where $G^{11}(\mathbf{r},\mathbf{r},\omega) = \sum_{s} G_{ss}(\mathbf{r},\mathbf{r},\omega)$ is the normal part of the
Nambu continuum Green’s function 
defined in a basis described by the field operators $\psi_s(\mathbf{r})$. These are related to the lattice operators $c_{\mathbf{R}ms}$ through the Wannier functions matrix elements $w_{\mathbf{R}m}(\mathbf{r})$ as $\psi_s(\mathbf{r})=\sum_{\mathbf{R}m}c_{\mathbf{R}ms} w_{\mathbf{R}m}(\mathbf{r})$. Employing the Wannier basis transformation\citep{Kreisel_PRL_2015, Kreisel_PRB_2016}, we can obtain the continuum Green's function as:
\begin{equation}
    {G}_{ss'}(\mathbf{r},\mathbf{r^\prime},\omega) = \sum_{\mathbf{RR^\prime}mn}
    \hat{G}_{ms,ns'}(\mathbf{R},\mathbf{R^\prime},\omega) w_{\mathbf{R}m}(\mathbf{r}) w_{\mathbf{R^\prime}n}(\mathbf{r^\prime})
    \label{Eqn:continuumGreen}
\end{equation}
and thereafter, evaluate the cLDOS from Eq.~(\ref{Eqn:cLDOS}). A simple Fourier transform of the cLDOS $\rho(\textbf{r},\omega)$ gives the Bogoliubov quasiparticle interference maps that can be compared directly to experimental measurements.

\begin{figure}[ht!]
    \centering
    \includegraphics[width=0.9\linewidth]{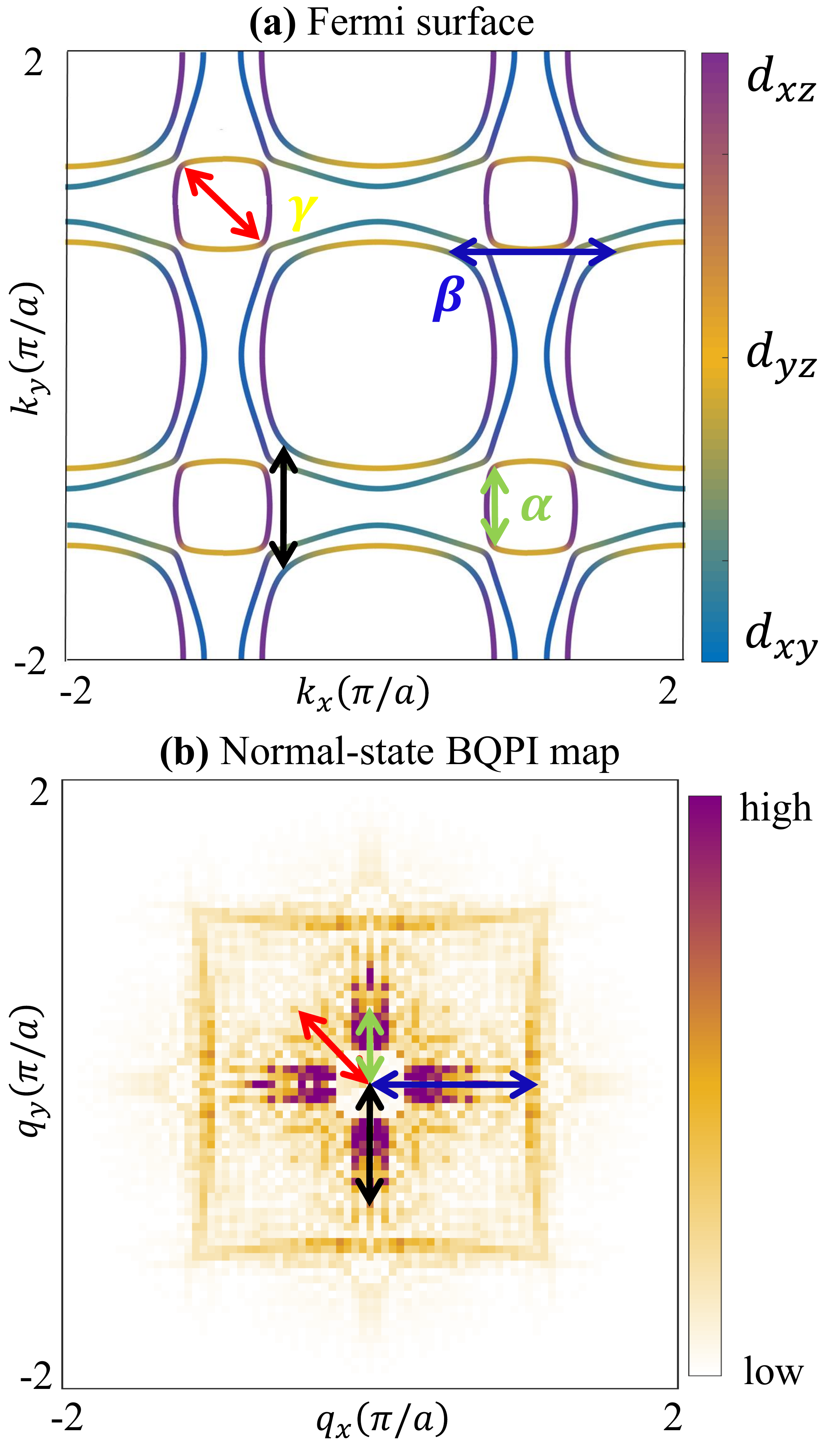}
	\caption{\textcolor{black}{(a) Fermi surface of {Sr}$_2${RuO}$_4$ for the model discussed in Sec.\ref{Subsec:non-int-H} with patches showing the dominant orbital content ($d_{xz}, d_{yz}$ or $d_{xy}$) as indicated by the color legend. The normal state intra-band scattering vectors connecting patches with similar orbital character are indicated by red, green, black and blue arrows. (b) The QPI pattern for an impurity potential $V_{\textnormal{imp}}= 0.05$ eV and for $\omega=0$ eV in the normal state of the system. The $\textbf{q}$-vectors corresponding to these red, green, black and blue arrows as shown in (a) are highlighted as well. Intensities in the central region surrounding $\textbf{q}=(0,0)$ have been suppressed {for visual clarity.}}}
    \label{Fig:SRO_fs}
\end{figure}

\section{Results} \label{Sec:results}

The {Sr}$_2${RuO}$_4$ crystal structure is composed of alternating layers of SrO and RuO$_2$ planes. The cleaving of the sample in ultrahigh vacuum at low temperatures is considered to reveal atomically flat SrO cleaved surface \citep{Madhavan_PNAS_2020}. However, an STM tip probing this surface is believed to be sensitive to the atomic wave functions of the Ru $d-$orbitals that dominate the Fermi level, while the other atomic wave functions away from the Fermi level are effectively invisible
for STM. An illustration of the Ru $d-$orbitals Wannier functions [as obtained from {\it ab-initio } calculations described in Appendix \ref{Appendix:wannier}] is provided in Fig.~\ref{Fig:SROwannier}. Notice the smaller $z$-expanse of the $d_{xy}$ Wannier orbitals compared to $ d_{xz},d_{yz}$ orbitals, making $d_{xy}$ orbital less likely to participate in tunneling through an STM tip located at a certain $z-$height above the cleaved surface.  This property might be altered on reconstructed surfaces\citep{kreisel2021unveiling}; in this work we concentrate on the spectroscopic features at low energies assuming no reconstruction  and  a negligible effect from the van Hove singularities.  Whether these effects are essential for the observation of superconductivity in Sr$_2$RuO$_4$ is not presently clear, as the presence or absence of reconstruction is not always examined in the experimental data\cite{Firmo2013, Madhavan_PNAS_2020}.

\begin{figure*}[hbt!]
    \centering
    \includegraphics[width=\textwidth]{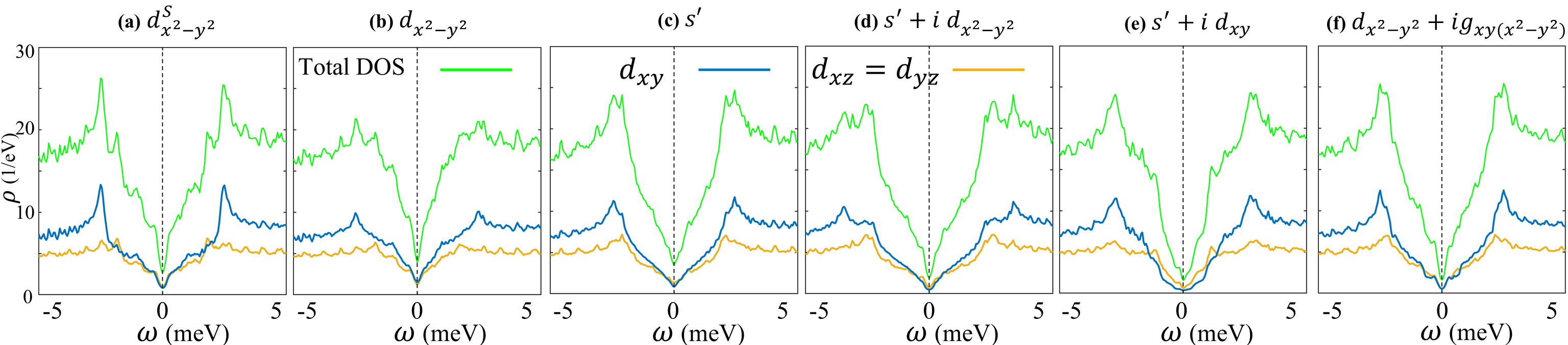}
	\caption{Homogeneous orbitally-resolved DOS in the superconducting state for six different gap order parameters: (a) $d^S_{x^2-y^2}$ which is a simple intra-band order parameter of $ \frac{\Delta_0}{2}(\cos \, k_x - \cos \, k_y)$, and the rest taken from Refs.~\onlinecite{Astrid_PRL_2019, Romer2021} (b) $d_{x^2-y^2} $, (c) $s' $, (d) $s'+id_{x^2-y^2} $, (e) $s'+id_{xy} $ and (f) $d_{x^2-y^2}+ig_{xy(x^2-y^2)} $. Notice the V-shaped DOS spectrum across all cases.}
    \label{Fig:DOS}
\end{figure*}

\subsection{Homogeneous superconducting state} \label{Subsec:HomoSC_gaps}

Figure~\ref{Fig:SRO_fs}(a) shows the Fermi surface (FS) for 2D {Sr}$_2${RuO}$_4$ with the corresponding dominant orbital contribution. There are three FS pockets: two quasi-1D bands that originate from Ru $d_{xz}$ (purple) and $d_{yz}$ (yellow) orbitals, leading to the electron-like $\beta$ band centered at the $\Gamma$ point and hole-like $\alpha$ band surrounding the M point; and the Ru $d_{xy}$ (blue) orbital generates the electron-like quasi-2D $\gamma$ band centered at the $\Gamma$ point. SOC lifts degeneracy of the $\beta$ and $\gamma$ bands along the diagonals in the first Brillouin zone (BZ) and introduces further inter-orbital hybridization as seen in Fig.~\ref{Fig:SRO_fs}(a). 

Figure~\ref{Fig:DOS} shows the orbitally-resolved homogeneous DOS $\rho^n(\omega)$ in the superconducting state, with  degenerate contributions from $d_{xz}$ and $d_{yz}$ orbitals  depicted by the dark-yellow line and $d_{xy}$ orbital contribution in blue. Results are presented for six different pseudospin-singlet gap order parameters: (a) purely intra-band $d^S_{x^2-y^2} = \frac{\Delta_0}{2}(\cos \, k_x - \cos \, k_y)$ {(the superscript `S' denotes the simple cosine form of the order parameter)}, and the rest adapted from Refs.~\onlinecite{Astrid_PRL_2019, Romer2021} (b) $d_{x^2-y^2} $, (c) $s' $, (d) $s'+id_{x^2-y^2} $, (e) $s'+id_{xy} $ and (f) $d_{x^2-y^2}+ig_{xy(x^2-y^2)} $. While the gap functions (b)-(f) correspond to specific cases of microscopic parameters discussed in those references, it was shown there that their structure was reasonably robust against changes of those parameters within reasonable ranges. Thus we hope therefore to identify qualitative BQPI structures that are driven by these features. Case (a) with $d^S_{x^2-y^2}$ order parameter was evaluated to compare directly with the BQPI analysis claims of Ref.~\onlinecite{Madhavan_PNAS_2020}. The gap maximum value for all cases was chosen as $\Delta_0 = 3.5$ meV, as required with the \textbf{k}-grid size employed.  A lower value of $\Delta_0$ would require a larger \textbf{k}-grid size for converged calculations; we verified that the qualitative features of computed BQPI patterns were robust against changes in the ${\bf k}$-grid.

The gap functions studied here all possess nodes or extremely deep near-nodes on the Fermi surface\cite{Astrid_PRL_2019,Romer2021}, such that  spectra in Fig. \ref{Fig:DOS} are all roughly V-shaped, consistent with STM measurements\cite{Kivelson_PRB_2013,Madhavan_PNAS_2020}. Certain small features related to the multiband character of {Sr}$_2${RuO}$_4$ are clearly visible, however. 
One can clearly identify the coherence peaks appearing in the vicinity of $\Delta_0 = \pm 3.5$ meV. Although some cases, particularly  (e) with $s'+id_{xy} $ order parameter,  show evidence of shoulders in the DOS at energies less than $\Delta_0$, this is not captured in experiments possibly due to: (1) the extremely low STM-bias resolution required to differentiate such a feature occurring below $\pm 350 \, \mu$eV in the real system, and (2) smoothing of the DOS spectrum due to convolution with Fermi distribution function at finite temperatures as measured in the $dI/dV$ spectrum \citep{Hoffman2011}.
We have used $1000\times 1000$ \textbf{k}-grid and a broadening parameter of $0.1$ meV for the above calculations.

\subsection{Inhomogeneous superconducting state}
\label{Subsec:inhomo_sc}

\begin{figure*}[h!]
  \centering    \includegraphics[scale=0.49,angle=0]{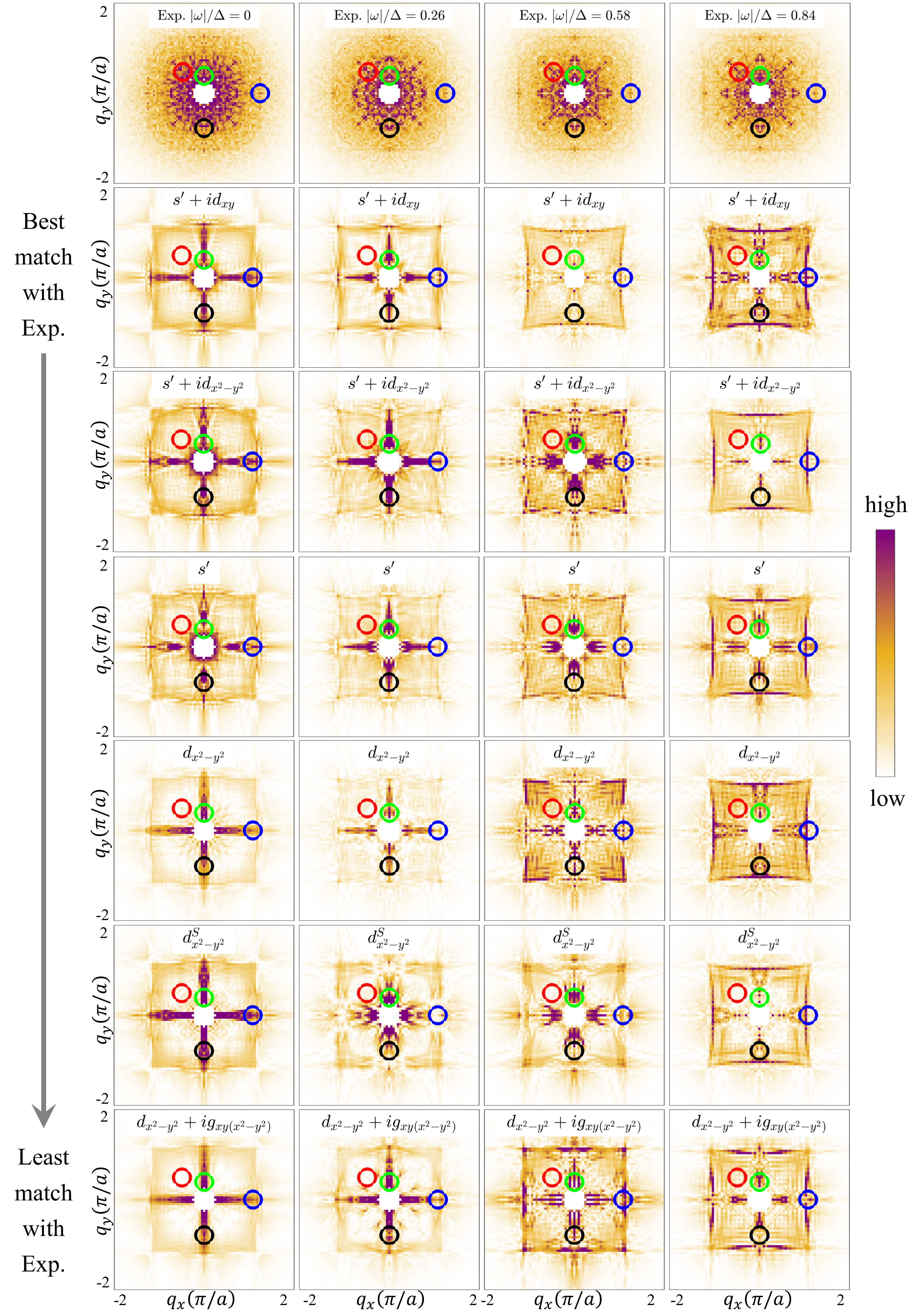}
    \caption{\textcolor{black}{BQPI maps obtained in the superconducting state from: (top row)  experimental data\cite{Madhavan_PNAS_2020} obtained at four different STM bias values $|\omega|/\Delta = [0,0.26,0.58,0.84]$ as marked for each column, followed by our theoretical predictions for $s'+id_{xy}$, $s'+id_{x^2-y^2}$, $s'$, $d_{x^2-y^2}$, $d^S_{x^2-y^2}$ and $d_{x^2-y^2}+ig_{xy(x^2-y^2)}$. The most noticeable $\textbf{q}$-features across all bias values are encircled in red, green, blue and black.
    Intensities in the region surrounding $\textbf{q}=(0,0)$ have been suppressed {for visual clarity}.}}
    \label{Fig:all_BQPI}
\end{figure*}

To consider scattering from a single point-like impurity, we introduce an impurity substituting one of the atoms in the center of the cleaved surface and calculate the cLDOS over an area of $51\times51$ lattice constants. We consider a weak non-magnetic impurity scatterer (Born limit) that is purely intra-orbital, and diagonal in spin space, i.e. $(V_{\textnormal{imp}}^{nn})_{ss} = 0.05$ eV appearing in Eq.~(\ref{Eqn:Tmatrix}).
Modeling the disorder is the most uncertain part of this analysis, since we do not have microscopic knowledge of the sources of scattering. We therefore follow the simplest path by modeling impurities by a single $\delta$-function with a potential chosen to best reproduce the simpler normal state QPI pattern. {Thus, here we do not consider various forms of dressed impurity potentials arising from electronic correlations\cite{Tsuchiura2001,ZWang2002,Zhu2002,Chen2004,Andersen2007,Harter2007,Andersen2007,Andersen2010,Schmid_2010,Gastiasoro2013,Sigrist_2021}. We evaluate the Wannier} function-modified cLDOS pattern obtained from Eq.~(\ref{Eqn:cLDOS}), first, for the normal state and next, for the superconducting states discussed under cases (a-f) in Sec.\ref{Subsec:HomoSC_gaps}. The spatial cLDOS is calculated on the $xy$-plane at a specific $z$-height above the cleaved surface of the sample (here, $z=4.93$\AA \,is used for the Wannier functions).
The specific choice of $V_{\textnormal{imp}}$ produces the QPI map of the normal state scattering processes at $\omega=0$ eV as shown in Fig.~\ref{Fig:SRO_fs}(b) and shows qualitative agreement with experimental results of Ref.[\onlinecite{Wang_2017,Madhavan_PNAS_2020}]. {The normal state intra-band scattering vectors connecting FS patches with similar orbital character across neighboring BZ are indicated by the red, green, black and blue arrows in Fig.~\ref{Fig:SRO_fs}(a). These specific $\textbf{q}$-vectors are highlighted in Fig.~\ref{Fig:SRO_fs}(b) representing the normal state QPI map. The specific double-cross feature was also observed in earlier experimental work\cite{Wang_2017,Madhavan_PNAS_2020}, but an interpretation of its origin was not discussed in those works. We found that most of the prominent \textbf{q}-features, including the sharp double-cross feature, arise from inter and intra-band $d_{xz/yz}$-scattering, as represented by the green, black and blue arrows. One can notice less-intense $\textbf{q}$-features corresponding to the suppressed tunneling matrix elements for $d_{xy}$ orbitals along the diagonals of the BZ indicated by the red arrow (which has been shown separately in Fig.~\ref{Fig:orbital_QPI}(b)). We have provided our detailed discussion about the $d_{xz/yz/xy}$-orbital contributions, as evidenced by the orbitally resolved normal state QPI maps, in Appendix \ref{Appendix:Orbital_QPI}.}

Next, we discuss the results for BQPI maps obtained in the superconducting state as shown in Fig.~\ref{Fig:all_BQPI} for cases (a-f). The top row shows the  experimental BQPI data\cite{Madhavan_PNAS_2020} obtained at four different STM bias values $|\omega|$ expressed in units of the experimentally observed gap-maxima $\Delta = 350 \, \mu$eV. For subsequent rows, $|\omega|$ is expressed in units of theoretically observed coherence peak values. These values correspond to: $|\omega|/\Delta = [0,0.26,0.58,0.84]$. The most consistently noticeable $\textbf{q}(\omega)$-features across all bias values are encircled in red, green, blue and black circles. Small $|\mathbf{q}|$-features in the experiments are believed to be due to long-range disorder/drift in real space rather than any scattering interference, and hence, are removed from the plots. We compared these specific features appearing in our theoretically evaluated BQPI maps corresponding to the six different singlet and composite order parameters.

Before proceeding, we note one salient aspect of the application of the BQPI technique to SRO and other low-$T_\text{c}$ unconventional superconductors.  While we have emphasized the utility of the technique to provide information on the gap structure in superconductors with low $T_\text{c}$, such systems often have much larger Dirac cone anisotropies $v_F/v_\Delta\sim E_F/\Delta_0$ than, e.g., cuprates or Fe-based superconductors.  As can be seen in the experimental data, there is therefore virtually no dispersion of the BQPI peak positions in the superconducting state, although the weights of these peaks change with bias.  Colloquially, this is because the contours of constant quasiparticle energy are arcs rather than ``bananas".  The dispersion takes place over a much smaller, and probably un-resolvable range of ${\bf q}$.  An analysis based on an analog of the ``octet" model applied to cuprates\cite{McElroy2003} is therefore not possible, eliminating one of the most powerful tools to identify the gap structure from BQPI.

We are therefore limited to comparing theoretical and experimental QPI maps at different biases and attempting to identify the most qualitatively robust features. 
In Fig.~\ref{Fig:all_BQPI}, we arranged the  rows of BQPI maps below the experimental data (first row) in a descending order starting from the case showing the most to the least consistency with the experimental patterns, i.e., second row for $s'+id_{xy}$, third for $s'+id_{x^2-y^2}$, fourth for $d^S_{x^2-y^2} $, fifth for $s' $, sixth for $d_{x^2-y^2}$ and seventh for $d_{x^2-y^2}+ig_{xy(x^2-y^2)}$.{ In other words, the BQPI pattern for $s'+id_{xy} $ order parameter shows the closest similarity at the four \textbf{q}-positions of the colored circles across all bias values, whereas $d_{x^2-y^2}+ig_{xy(x^2-y^2)} $ order parameter shows the least correspondence with these four specific \textbf{q}-features. These four \textbf{q}-positions were first identified from their ubiquitous and robust presence across all bias values in the experimental BQPI maps, and thereafter, their corresponding presence in the theoretically predicted BQPI maps were investigated \cite{Madhavan_PNAS_2020}.  Furthermore, to quantify the similarities, we evaluated the mean-squared deviation (MSD) $ \overline{\Delta y(j)}$ of the integrated intensities around the four specific \textbf{q}-points across all STM bias values for different order parameters $(j)$, weighted inversely by their extent of mismatch  compared to their experimental counterparts. A summary score chart of this correspondence is quantified in Table \ref{table:1} below and a more detailed description is given in Appendix \ref{Appendix:Scorechart}. Additionally, the experimental features at higher STM biases $|\omega|/\Delta = [0.58,0.84]$ show enhanced intensities along the square edges positioned around $(\pm 1.5 \pi/a)$, as opposed to the lower biases $|\omega|/\Delta = [0,0.26]$. This trend is also observed in the BQPI maps for the $s'+id_{xy} $ and the $s'+id_{x^2-y^2} $ order parameters, whereas the other order parameters tend to exhibit this feature at other bias values, not aligning with this trend. This enhanced square feature is believed to be arising from increased scattering between gradually re-appearing normal-state-like Bogoliubov contours at higher bias values.
}

\begin{table}[b!]
\centering
\begin{tabularx}{1.\linewidth} { 
  | >{\centering\arraybackslash}X 
  | >{\centering\arraybackslash}X | }
\hline
Gap order & Weighted Mean Squared Deviation 
\\
\hline
\hline
$s'+id_{xy} $ & 2.79  \\
\hline
$s'+id_{x^2-y^2} $ & 2.93  \\
\hline
$d^S_{x^2-y^2} $ & 2.94 \\
\hline
$s' $ & 3.07 \\
\hline
$d_{x^2-y^2} $ & 3.17 \\
\hline
$d_{x^2-y^2}+i g_{xy(x^2-y^2)} $ & 5.62 \\
\hline 
\end{tabularx}
\caption{\textcolor{black}{Score chart summarizing the weighted mean-squared deviation $ \overline{\Delta y(j)}$ of the integrated BQPI intensity around the four specific \textbf{q}-features at the colored circles (see Fig.~\ref{Fig:all_BQPI}) between experimental and theoretically evaluated BQPI maps, for various gap order parameters $(j)$. The numbers are sorted starting from the least $(s'+id_{xy})$ to the highest $(d_{x^2-y^2}+ig_{xy(x^2-y^2)})$ MSD corresponding to various order parameters. This score determines the ranking for the correspondence of the individual gap order parameters when matched with experimental BQPI patterns.}}
\label{table:1}
\end{table}

{As discussed in Ref.~\onlinecite{Madhavan_PNAS_2020}, the $d_{xz/yz}$ dominated $\alpha,\beta$ bands are detected from the normal-state scattering interference wave-vectors (blue, green and black arrows in Fig. \ref{Fig:SRO_fs}(a)) and subsequently, yield prominent signatures in the superconducting state.
With the knowledge that our gap order parameter exhibits multiple nodes and/or minima in the vicinity of  $(\pm\pi/a,\pm\pi/a)$ on $\alpha,\beta$ bands, there could be numerous scattering wave-vectors connecting these $\mathbf{k}$-regions. This gives rise to the rich and intricate BQPI patterns as observed in experiment and in our theoretical evaluations as well. However, we note that the most significant features highlighted by the blue, green and black circles (in Fig.~\ref{Fig:all_BQPI}) correspond to major scattering wave-vectors connecting the Bogoliubov contours along the $(\pm\pi/a,\pm\pi/a)$ on $\alpha$ and $\beta$ bands. Detecting BQPI intensities along the diagonal regions of the BZ is somewhat challenging in our numerical evaluations. This is because, in our current tight-binding model { together with} \textit{ab-initio} derived Wannier functions, the small value of the $d_{xy}$ Wannier function significantly suppresses the scattering matrix elements. Further details are discussed in Appendix \ref{Appendix:Orbital_QPI}.}

\section{Summary} \label{Sec:summary}

{The question of the gap structure of Sr$_2$RuO$_4$ in momentum space is difficult to address experimentally due to the small superconducting energy scales in this system.  At present, Bogoliubov quasiparticle interference measurements appear to be the only method of pinning down the location of gap nodes.}
We have{ therefore} performed 2D calculations of the real-space Wannier function-modified continuum electronic density of states for {Sr}$_2${RuO}$_4$. Using the T-matrix approach in combination with Wannier functions for analyzing scattering effects from a single impurity, we evaluated the real-space tunneling conductance map and displayed its Fourier transformed BQPI image to compare with STM measurements \citep{Madhavan_PNAS_2020}. 
We have shown our evaluations for six possible pseudospin-singlet and composite gap order parameters: (a) purely intra-band $d^S_{x^2-y^2}$ parameter  $( \frac{\Delta_0}{2}(\cos \, k_x - \cos \, k_y))$, and the rest borrowed from Refs.~\onlinecite{Astrid_PRL_2019, Romer2021}: (b) $d_{x^2-y^2} $, (c) $s' $, (d) $s'+id_{x^2-y^2} $, (e) $s'+id_{xy} $, and (f) $d_{x^2-y^2}+i g_{xy(x^2-y^2)} $. We compared the important features between our theoretical predictions and the experimental measurements for the homogeneous DOS spectrum and the BQPI patterns, and concluded that the $s'+id_{xy} $ gap order parameter seems to be the most consistent with our observations. {However, the analysis is at best quantitative;
the overall agreement with the measured patterns does not seem sufficiently impressive at this time to make a strong case for any of the above 2D states.}
Since the analysis of BQPI patterns is a classic example of ``pattern recognition'' problem\citep{Zhang2019_EQM}, future investigations in this context can be directed towards training convolutional neural networks with theoretically generated images of BQPI patterns labeled for various STM biases and gap order parameters, and testing them on experimental data sets. This will facilitate less manual interference in analyzing complicated and rich patterns in data, and will be less prone to bias and estimation errors.

\begin{acknowledgments}
    We thank Rahul Sharma and J.C. Davis for sharing experimental data with us and for many helpful discussions that paved the course of the analysis presented in this work. We also thank  Ilya Eremin, Marvin M\"uller, Miguel Sulangi, Roser Valent{\'i} and Karim Zantout for useful discussions. S.B.  acknowledges support from the Deutsche Forschungsgemeinschaft (DFG, German Research Foundation) through TR 288-422213477. B. M. A. and A. T. R. acknowledge support from the Independent Research Fund Denmark grant number 8021-00047B. P.J.H. and S.B. acknowledge partial support from DOE-BES DE-FG02-05ER46236. A portion of this research was conducted at the Center for Nanophase Materials Sciences, which is a DOE Office of Science User Facility (T. B., X. R. K.). This research used resources of the Compute and Data Environment for Science (CADES) at the Oak Ridge National Laboratory, which is supported by the Office of Science of the U.S. Department of Energy under Contract No. DEAC05-00OR22725.
\end{acknowledgments}

\appendix

\section{First principles Wannier function calculations}
\label{Appendix:wannier}

\begin{figure}[ht!]
    \centering
    \includegraphics[width=\linewidth]{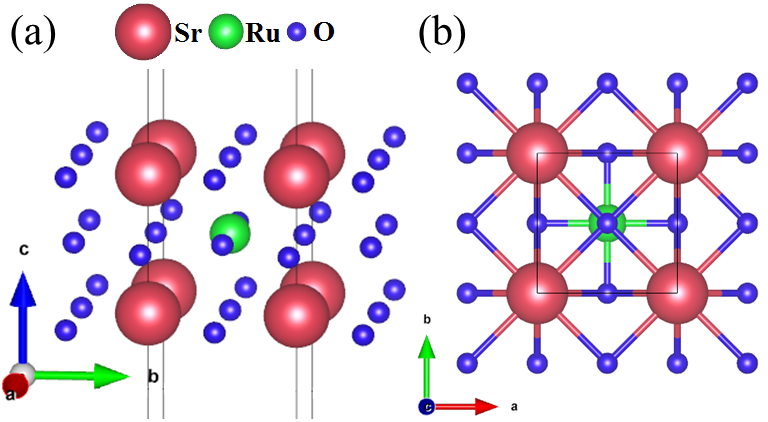}
	\caption{(a) The cell (constituent atoms and vacuum) of {Sr}$_2${RuO}$_4$ depicted with black borders and constituent atoms Sr (maroon), Ru (green) and O (blue) as indicated by the color legend. The upper and lower extension of the cell into vacuum has not been shown for brevity. Axis orientations are shown in the left-bottom corner. (b)  Surface geometry of {Sr}$_2${RuO}$_4$ when viewed above a certain height from the surface, with the Sr (maroon) and O (blue) atoms on very similar $z$-level. The bonds (in-plane and out-of-plane) between different atoms are depicted by the bi-colored lines.}
    \label{Fig:SRO_DFT}
\end{figure}

The Wannier functions in this work were derived following the same procedure as in Ref.~\onlinecite{kreisel2021unveiling}. In this work however, no octahedral rotations were incorporated. Therefore, we could use the {Sr}$_2${RuO}$_4$ normal cell instead of a $\sqrt{2}\times\sqrt{2}$ {Sr}$_4${Ru}$_2${O}$_8$ supercell. As a first step, we performed Density Functional Theory (DFT) calculations of monolayer {Sr}$_2${RuO}$_4$ with an in-plane lattice constant $a = 3.8644$ \AA{} and a vacuum layer of roughly 20 \AA. The corresponding unit cell is depicted in Fig.~\ref{Fig:SRO_DFT}. Including the vacuum in our calculations allowed us to simulate the Wannier functions above the surface of {Sr}$_2${RuO}$_4$ where the STM tip resides. For the DFT calculations we employed the Vienna ab initio simulation package (VASP) \citep{Kresse1996,Kresse1999}. We used the generalized gradient approximation of Perdew, Burke and Ernzerhof \citep{PBE1996}, a plane-wave energy cut-off of $650$ eV and a $7 \times 7 \times 1$ k-mesh. In the second step, we derived the Wannier functions with the Wannier90 software \citep{Mostofi_2008}. Specifically, we projected the Ru-$t_{2g}$ orbitals on the bands roughly within $[-9.55,1.45]$ eV. Furthermore, we set $\mathtt{num\_iter}=0$ and $\mathtt{dis\_num\_iter}=10000$, and took the inner frozen energy window to be roughly equal to $[-1.65,1.45]$ eV.

\section{Representing pairing functions in orbitally-resolved Nambu basis}
\label{Appendix:Spatialgap}

\begin{figure*}[hbt!]
    \centering
    \includegraphics[width=\textwidth]{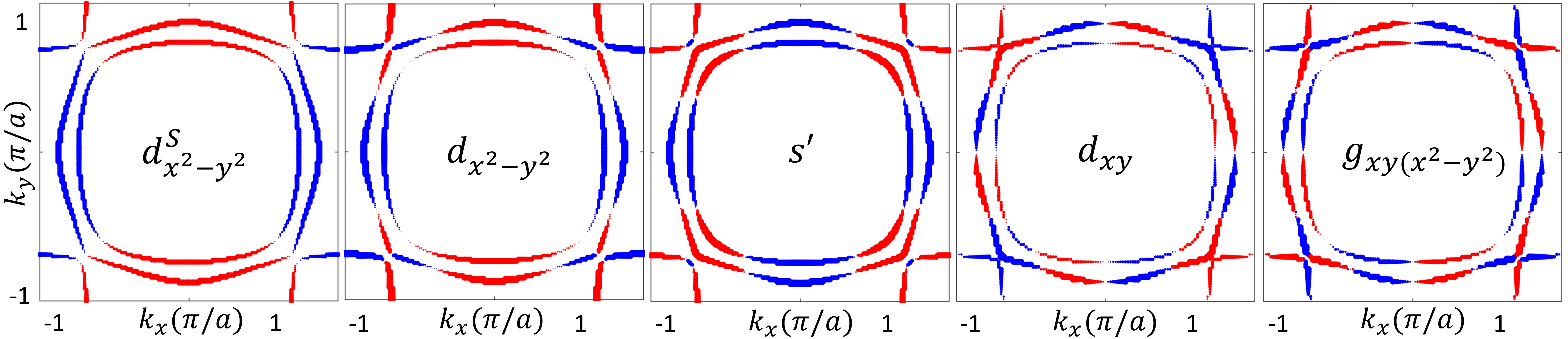}
	\caption{Depictions of gap order parameters as a function of the Fermi surface for $d^S_{x^2-y^2}$ order parameter (simple intra-band form-factor of $ \frac{\Delta_0}{2}(\cos \, k_x - \cos \, k_y)$), and the rest borrowed from Refs.~\onlinecite{Astrid_PRL_2019, Romer2021} $d_{x^2-y^2} $, $s' $, $d_{xy} $ and $g_{xy(x^2-y^2)} $. The red (blue) color denotes positive (negative) values of the order parameter and marker-size of each $\mathbf{k_F}$ point is proportional to the gap magnitude $|\Delta(\mathbf{k_F})|$. Composite order parameters like $s'+id_{x^2-y^2} $, $s'+id_{xy} $ and $d_{x^2-y^2} + ig_{xy(x^2-y^2)}$ were obtained from complex combinations of the corresponding individual order parameters.}
    \label{Fig:gaps}
\end{figure*}

The gap function obtained from Refs.~\onlinecite{Astrid_PRL_2019, Romer2021} and illustrated in Fig.~\ref{Fig:gaps} are expressed as a function of the normal state Fermi surface (FS) $\Delta_\mu(\mathbf{k_F})$, where the gap magnitude tends to be the largest compared to the rest of the BZ. To obtain the gap function over the full BZ as required in Eq.~(\ref{Eqn:fullBCS}), one can extrapolate its value from the $\mu$-th Fermi band $\Delta_\mu(\mathbf{k_F})$ as it falls off away from the FS, a behavior which can be parametrized in terms of a Gaussian-cutoff:
\begin{align}  
   \Delta_\mu(\mathbf{k}) = \Delta_\mu( \mathbf{k_F}^{\textnormal{nn.}}) \; \text{exp}\left(- \left( |E_\mu(\mathbf{k})|/\Delta_C\right)^2\right),
   \label{Eq:gap_extrapolation}
\end{align}
where $\Delta_C = 10\Delta_0$ and $\mathbf{k_F}^{\textnormal{nn.}}$ refers to nearest-neighbor $\mathbf{k_F}$ from $\mathbf{k}$. This provides a local picture of the internal orbital structure of a pair which continues out to a radius set by the coherence length $\xi_0$. While implementing this extrapolation, one should keep in mind the translational invariance condition for the BZ, i.e, to obtain $\Delta_\mu(\mathbf{k})$ values in the 1$^{st}$ BZ ($-\pi/a:\pi/a$), contributions of $\Delta_\mu(\mathbf{k_F})$ values from the neighboring BZs ($-2\pi/a:2\pi/a$) should also be taken into account. The gap maxima value of $|\Delta_\mu(\mathbf{k})|$ in the BZ can also be set to a chosen $\Delta_{0}$, for instance, 3.5 meV for our numerical purpose. The following describes the next steps of basis transformation from bands to orbital basis, for representing the gap (pairing) functions. 

\subsection{Tight-binding Hamiltonian}

The non-interacting Hamiltonian $\hat{H}(\textbf{k})$ as in Eq.~(\ref{Eqn:H0}) and (\ref{Eqn:HSOC}), has the block diagonal structure:
\begin{equation}
   \hat{H}(\mathbf{k}) = \left( \begin{array}{cc} 
    H_0(\mathbf{k}) + H_{\textnormal{soc}}(+) & 0 \\
    0 & H_0(\mathbf{k}) + H_{\textnormal{soc}} (-)
    \end{array} 
    \right),
\end{equation}
Since $H_{\textnormal{soc}}(\sigma=\pm)$ has imaginary components, diagonalization of $\hat{H}(\mathbf{k})$ can yield eigenvectors that are linear combination of Kramer's degenerate eigenstates owing to gauge-independence. To circumvent this, one can diagonalize $H_+(\mathbf{k}) = H_0(\mathbf{k}) + H_{\textnormal{soc}}(+)$, and obtain the unitary matrix $U^{(+)}(\mathbf{k})$ and then $H_-(\mathbf{k}) = H_0(\mathbf{k}) + H_{\textnormal{soc}}(-)$, to obtain $U^{(-)}(\mathbf{k})$, and form the unitary matrix $U^{(0)}(\mathbf{k})$ such that:
\begin{align}
    \label{Eqn:unitary_matrix}
    & U^{(0)}(\mathbf{k}) = \left( \begin{array}{cc} 
    U^{(+)}(\mathbf{k}) & 0 \\
    0 & U^{(-)}(\mathbf{k})
    \end{array} 
    \right), \\
    & U^{(0)\dagger}(\mathbf{k}) \hat{H}(\mathbf{k}) U^{(0)}(\mathbf{k}) = \left( \begin{array}{cc} 
    E^{(+)}(\mathbf{k}) & 0 \\
    0 & E^{(-)}(\mathbf{k})
    \end{array} 
    \right), 
\end{align}
$U^{(0)}(\mathbf{k})$ diagonalizes $ \hat{H}(\mathbf{k})$ to yield Kramer's degenerate pairs of eigenvalues $E^{(+)}(\mathbf{k})=E^{(-)}(\mathbf{k})$. 

\subsection{Pairing function}

The gap part of the BCS Hamiltonian $\hat{\Delta}(\mathbf{k})$ is constructed in the same basis as the $\hat{H}(\mathbf{k})$. However, the gap values obtained from Eq.~(\ref{Eq:gap_extrapolation}) are represented in band pseudospin basis. For the three bands present in the {Sr}$_2${RuO}$_4$ system and for{ pseudospin-singlet} pairing with the Cooper-pair structure:  $|\uparrow \downarrow \rangle - | \downarrow \uparrow \rangle$, we construct the gap matrix in this band pseudospin basis as $    \Delta^{\textnormal{pseudo}}(\mathbf{k}) $:
\begin{equation}
 \left( \begin{array}{cccccc} 
     . & . & . & \Delta_1(\mathbf{k}) & . & . \\
     . & . & . & . & \Delta_2(\mathbf{k}) & . \\
     . & . & . & . & . & -\Delta_3(\mathbf{k}) \\
     -\Delta_1(\mathbf{k}) & . & . & . & . & . \\
     . & -\Delta_2(\mathbf{k}) & . & . & . & . \\
     . & . & \Delta_3(\mathbf{k}) & . & . & . \\
    \end{array} 
    \right), 
\end{equation}
where the pseudospin-structure of the basis is $\left( \uparrow , \uparrow, \downarrow,\downarrow,\downarrow,\uparrow \right)$.
One constructs the gap part of the Hamiltonian in the Nambu basis as
\begin{align}
    \underline{\hat{H}}_\Delta &=     \left( \begin{array}{cc} 
    0 & \Delta^{\textnormal{pseudo}}(\mathbf{k}) \\
    \left(\Delta^{\textnormal{pseudo}}(\mathbf{k}) \right)^\dagger & 0
    \end{array} \right ).
\end{align}
The unitary matrix in this Nambu basis \begin{align}
   \underline{U}_N(\mathbf{k}) &= \left( \begin{array}{cc} 
    U^{(0)}(\mathbf{k}) & 0 \\
    0 & \left(U^{(0)}(-\mathbf{k})\right)^*
    \end{array} 
    \right),
\end{align}
that diagonalizes the non-interacting part of the Nambu Hamiltonian 
\begin{equation}
    \underline{\hat{H}}_N = \left( \begin{array}{cc} 
    \hat{H}(\mathbf{k}) & 0 \\
     0 & -\hat{H}^T(-\mathbf{k})
    \end{array} 
    \right),
\end{equation}
can be applied on $\underline{\hat{H}}_\Delta$ for unitary transformation from the band to orbital basis:
\begin{equation}
    \underline{U}_N(\mathbf{k}) \underline{\hat{H}}_\Delta(\mathbf{k}) \underline{U}^\dagger_N(\mathbf{k}) = \left( \begin{array}{cc} 
    0 & \hat{\Delta}(\mathbf{k}) \\
    \hat{\Delta}^\dagger(\mathbf{k}) & 0
    \end{array} 
    \right).
\end{equation}   
The matrix elements $\Delta_{mn}(\mathbf{k})$ of the block $\hat{\Delta}(\mathbf{k})$ are the orbitally resolved gap structure in the Nambu basis. Its real-space representation is obtained by a simple Fourier transformation:
\begin{align}  
  \Delta_{mn}(\mathbf{R}) &= \frac{1}{N_k}\sum_{\mathbf{k}} \Delta_{mn}(\mathbf{k}) e^{-i \mathbf{k.R}},
\end{align}
where $\mathbf{R}$ refers to different lattice sites centered around (0,0). The amplitude $\Delta_{mn}(\mathbf{R})$ encodes the internal spatial and orbital structures of the electron pair. With the information of $\Delta_{mn}(\mathbf{R})$ and its simple inverse Fourier transformation back to momentum space, one can set up the full BCS Hamiltonian of Eq.~(\ref{Eqn:fullBCS}) in combination with any tight-binding parameters.

\section{Orbital contribution to the normal state QPI patterns} \label{Appendix:Orbital_QPI} 

\begin{figure}[ht!]
    \centering
    \includegraphics[width=\linewidth]{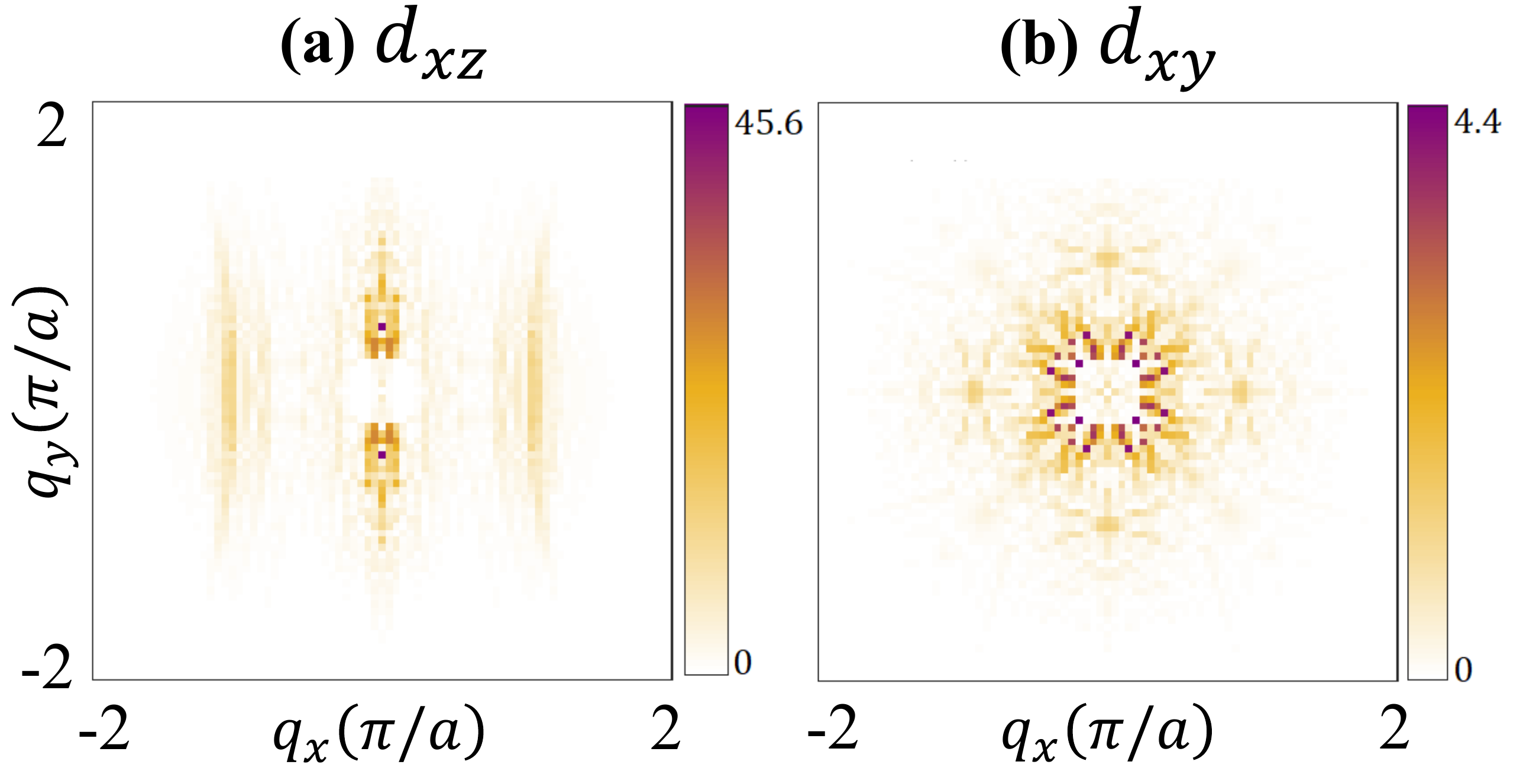}
	\caption{\textcolor{black}{Orbitally-resolved normal state QPI map for {Sr}$_2${RuO}$_4$ at $\omega=0$ for (a) $d_{xz}$ and (b) $d_{xy}$ orbitals. Notice that the QPI intensity magnitude are different for the two orbitals, with the $d_{xy}$ orbital intensity being $\sim 10$ times weaker than the $d_{xz}$ orbitals. The $d_{yz}$-orbital QPI map is same as in $d_{xz}$, but rotated by $ \pi/2$.  }}
    \label{Fig:orbital_QPI}
\end{figure}
\begin{figure}[ht!]
    \centering
    \includegraphics[width=0.99\linewidth]{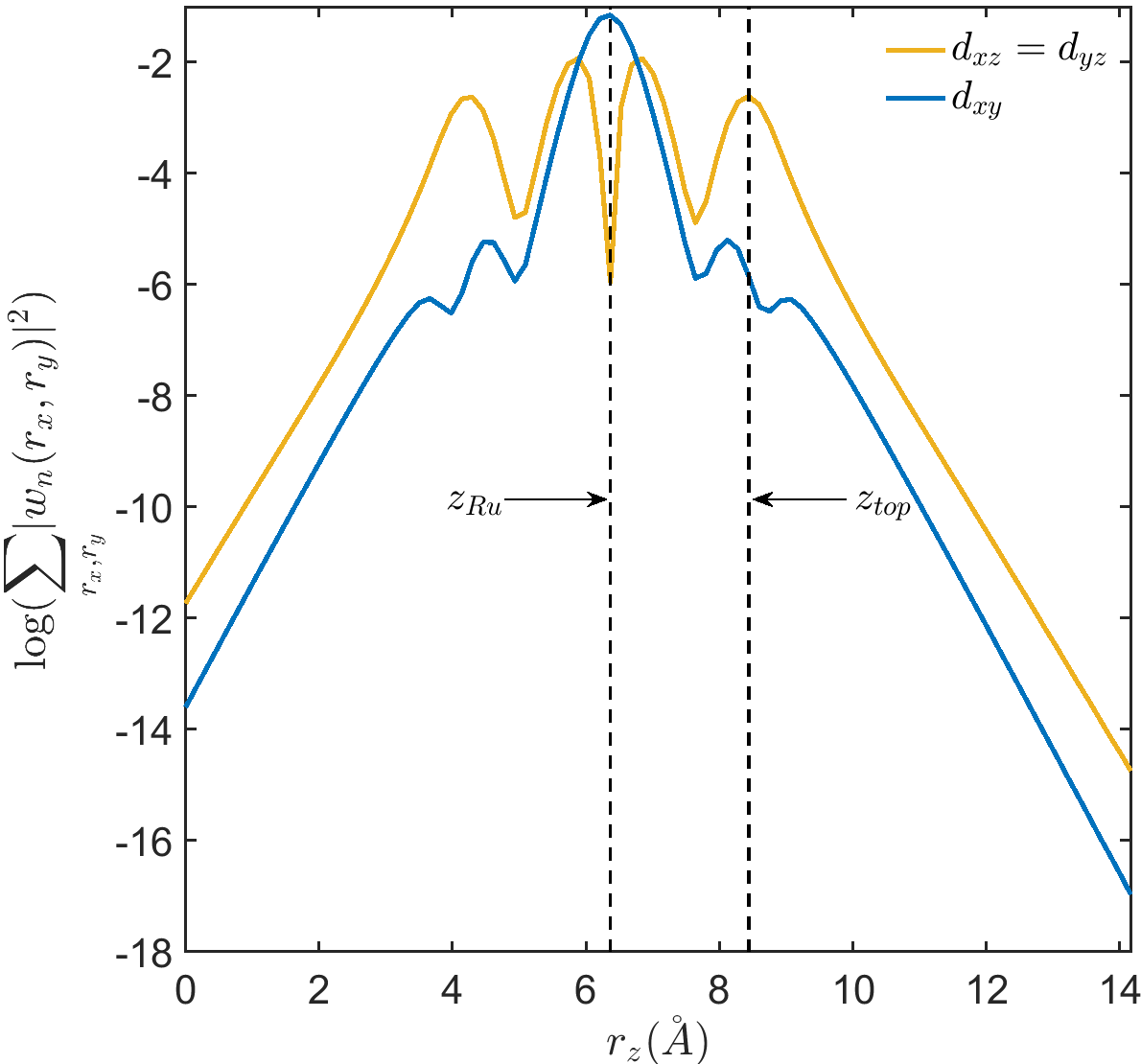}
	\caption{A log$_{10}$-plot of the orbitally-resolved $x,y$-integrated ruthenium Wannier wave function amplitude as a function of the $z$-value of the unit cell representing {Sr}$_2${RuO}$_4$. The center of the unit cell containing the Ru atom has been marked by $z_{Ru}$ in the figure. The top surface of the unit cell is denoted by $z_{top}$. The Wannier wave functions for the $d_{xz}(=d_{yz})$ orbitals (yellow line) expand beyond the unit cell along the $z$-direction with a larger magnitude compared to the restricted spread of the $d_{xy}$ orbitals (blue line).}
    \label{Fig:orbital_wannier_tail}
\end{figure}

\textcolor{black}{Here, we discuss the individual orbital contributions towards the normal state QPI pattern as shown in Fig.~\ref{Fig:orbital_QPI}, evaluated as per the following equation for $m$-th orbital:
\begin{equation}
    {\rho}_{m}(\mathbf{r},\omega) = -\frac{1}{\pi} \textrm{Im} \sum_{\mathbf{RR^\prime}s}
    \hat{G}_{ms,ms}(\mathbf{R},\mathbf{R^\prime},\omega) w_{\mathbf{R}m}(\mathbf{r}) w_{\mathbf{R^\prime}m}(\mathbf{r})
    \label{Eq:orbital_cldos}
\end{equation}
where $\hat{G}_{ms,ms}(\mathbf{R},\mathbf{R^\prime},\omega)$ is the lattice Green's function in the presence of impurities (Eq.~(\ref{Eqn:ImpurityGreen_Tmatrix})) and $ w_{\mathbf{R}m}(\mathbf{r})$ are the Wannier functions matrix elements.
First, we attribute the double-cross feature seen in Fig.~\ref{Fig:SRO_fs}(b) to intra- and inter-band scattering between the $d_{xz}/d_{yz}$-dominated Fermi pockets in neighboring BZ. This can be verified from the $d_{xz}$-resolved QPI map in Fig.~\ref{Fig:orbital_QPI}(a). These intraband scattering vectors have already been depicted in Fig.~\ref{Fig:SRO_fs}(a) by the green, blue and black arrows. We have not marked the interband scattering vectors, but they can be visualized as vectors connecting the $d_{xz}$-dominated parts of the $\beta$ and the $\alpha$ pockets, closely aligned along the black arrow in Fig.~\ref{Fig:SRO_fs}(a). The less-intense $\textbf{q}$-features along the diagonals of the BZ, correspond to the suppressed tunneling matrix elements for $d_{xy}$ orbitals as shown in Fig.~\ref{Fig:orbital_QPI}(b). One of these features is marked by the red arrow in Fig.~\ref{Fig:SRO_fs}(b). Notice that the  $d_{xz}(=d_{yz})$ features are $\sim 10$ times stronger than the $d_{xy}$ features. Note that the partial contribution to the density of states is given by the product of the lattice Green's function and the Wannier functions (see Eq.\ref{Eq:orbital_cldos}). To understand 
the difference in the intensities of the QPI maps, one needs to analyze the modulation of the lattice Green's function in conjunction with the Wannier wave functions. 
In Fig.~\ref{Fig:orbital_wannier_tail}, we show a log$_{10}$-plot of the orbitally-resolved $x,y$-integrated ruthenium Wannier wave function amplitude as a function of the $z$-value of the unit cell representing {Sr}$_2${RuO}$_4$. The integrated Wannier function amplitude is symmetric around the Ru atoms for both the orbitals, as expected. However, the $d_{xz}(=d_{yz})$ orbitals (yellow line) expand beyond the unit cell along the $z$-direction with a larger magnitude compared to the restricted spread of the $d_{xy}$ orbitals (blue line). At a distance of $4.93$\AA  \, above the surface, i.e. at $z=13.93$\AA, where we carry out our calculations, the $d_{xz}(=d_{yz})$ orbital integrated Wannier function amplitude is $\sim 10$ times stronger than the $d_{xy}$ orbital. This ratio is exactly what we observe in the orbitally-resolved QPI intensity differences in Fig.~\ref{Fig:orbital_QPI}. This difference between the orbital wave function intensity washes out most of the $d_{xy}$ features and enhances the $d_{xz/yz}$ features in our BQPI maps.} 

\section{Detailed description of the analysis of the BQPI patterns for ranking the different gap order parameters}
\label{Appendix:Scorechart}

\textcolor{black}{In this section, we describe the score assignment procedure for ranking the different order parameters as per their BQPI pattern matching with the experiments. 
As mentioned earlier, we identified four \textbf{q}-positions from their ubiquitous and robust presence across all STM bias values in the experimental BQPI maps and marked them in red, green, blue and black colored circles (see Fig.~\ref{Fig:all_BQPI}, first row)}.

\begin{figure*}[hbt!]
    \centering
    \includegraphics[width=\linewidth]{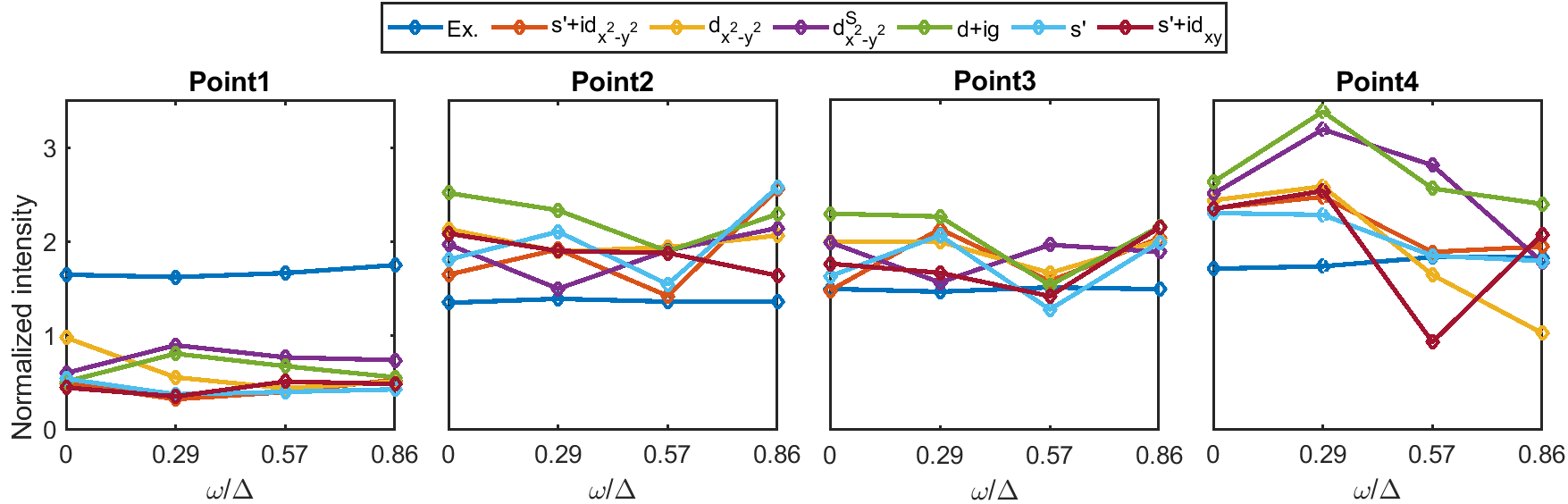}
	\caption{\textcolor{black}{Left to right: Normalized integrated intensities around each \textbf{q}-point, encircled in red (point 1), blue (point 2), green (point 3) and black (point 4) colors in Fig.~\ref{Fig:all_BQPI}, as a function of STM bias values $\omega/\Delta$, for different gap order parameters as indicated by the legend on top. The normalization was done with respect to the fully integrated BQPI intensity over the extended BZ at each bias value for each order parameter.}}
    \label{Fig:least_MSD}
\end{figure*}

{Next, we evaluate the integrated intensities of the BQPI maps around these four specific \textbf{q}-points, as a function of different STM bias values and for different order parameters along with the experimental data, as shown in Fig.~\ref{Fig:least_MSD}. This was done by selecting a patch of $15\times 15$ pixels centered at each colored circle. These intensities were normalized by the total integrated intensity over the full extended BZ at each STM bias for each order parameter. 
Next, we evaluate the mean-squared deviation (MSD) $ \Delta y(j,i)$ of the integrated BQPI intensity between experimental and numerically evaluated BQPI maps for the 4 specific \textbf{q}-features centered inside the colored circles $(i)$ for various gap order parameters $(j)$. Table \ref{table:2} quantifies the details of this analysis with rankings shown in descending order as before. The deviations are summed across all STM bias values. Since some \textbf{q}-points (example, red circle) tend to display a higher mean squared-deviation from the experimental counterpart than the other \textbf{q}-points, therefore, we assign a weight $ \textrm{w}(i)$ corresponding to each colored circle to reflect the nature of this discrepancy. To evaluate this weight $ \textrm{w}(i)$, first, we find the cumulative deviation corresponding to each colored circle across all order parameters. We pick the circle corresponding to the minimum of this cumulative deviation weight, i.e., the green point in this case. Next, we find normalized weights in units of this minimum cumulative weight by dividing the total weights by this minimum cumulative weight $= 5.13$ (written in bold in Table \ref{table:2}). When we re-evaluate the weighted MSD $ \overline{\Delta y(j)}$, we multiply the normalized weights $ \overline{\textrm{w}(i)}$ with the MSD $ \Delta y(j,i)$  corresponding to each circle $i$ for a specific order parameter $j$, to find the $ \overline{\Delta y(j)}$.  This $ \overline{\Delta y(j)}$ score determines the ranking for the correspondence of the individual gap order parameters when matched with experimental BQPI patterns:
\begin{align}
    \overline{\Delta y(j)}  = 
    \sum_{\textrm{point }{i=1,2,3,4}} \Delta y(j,i)  \; \; \overline{\textrm{w}(i)}
\end{align}
as provided in Table \ref{table:1} in the main text.
An additional experimental feature is the enhanced intensities along the square edges positioned around $(\pm 1.5 \pi/a)$ at higher STM biases $|\omega|/\Delta = [0.58,0.84]$ as opposed to the lower biases $|\omega|/\Delta = [0,0.26]$. We observe that the $s'+id_{xy}$ order parameter also displays a similar trend in its BQPI map, while other order parameters display different trends.
Our approach is based on the choice of dominant \textbf{q}-features and results might change if different choices were made. Unfortunately, we do not have a more robust way to approach this pattern-matching problem. We are restricted by the limited availability of experimental data at ultra-low temperatures. }

\begin{table}[th!]
\centering
\begin{tabularx}{1.\linewidth} { 
  | >{\centering\arraybackslash}X
  | >{\centering\arraybackslash}X 
  | >{\centering\arraybackslash}X 
  | >{\centering\arraybackslash}X 
  | >{\centering\arraybackslash}X | }
\hline
 \multicolumn{5}{|c|}{MSD $ \Delta y(j,i)$ from exp. intensity data}
 \\
\hline
Gap order & Red point & Blue point & Green point & Black point \\
\hline
\hline
 $s'+id_{x^2-y^2} $& 6.12 & 1.81 & 0.74 & 0.98 \\
\hline
$d_{x^2-y^2} $ & 4.63 & 1.70 & 0.82 & 1.94 \\
\hline
$d^S_{x^2-y^2} $ & 3.45 & 1.31 & 0.62 & 3.72 \\
\hline
$d_{x^2-y^2}+i g_{xy(x^2-y^2)} $ & 4.37 & 3.41 & 1.72 & 4.42 \\
\hline
$s' $ & 6.13 & 2.24 & 0.68 & 0.65 \\
\hline
$s'+id_{xy} $ & 5.99 & 1.15 & 0.55 & 1.93 \\
\hline 
\hline 
Total $ \textrm{w}(i)$ & 30.69 & 11.62 & \textbf{5.13} & 13.64 \\
\hline
Normalized $ \overline{\textrm{w}(i)}$ & 0.17 & 0.44 & 1.0 & 0.38 \\
\hline
\end{tabularx}
\caption{{Table displaying the mean-squared deviation $ \Delta y(j,i)$ of the integrated BQPI intensity between experimental and numerically evaluated BQPI maps, around the four specific \textbf{q}-features at the colored circles  $(i)$ (see Fig.~\ref{Fig:all_BQPI}) for various gap order parameters $(j)$. The last two rows show the summed total weights $ \textrm{w}(i)$ and the normalized weights $ \overline{\textrm{w}(i)}$, as described in the text above in Appendix \ref{Appendix:Scorechart}.}}
\label{table:2}
\end{table}

\bibliography{references_BQPI}{}

\begin{thebibliography}{73}%
\makeatletter
\providecommand \@ifxundefined [1]{%
 \@ifx{#1\undefined}
}%
\providecommand \@ifnum [1]{%
 \ifnum #1\expandafter \@firstoftwo
 \else \expandafter \@secondoftwo
 \fi
}%
\providecommand \@ifx [1]{%
 \ifx #1\expandafter \@firstoftwo
 \else \expandafter \@secondoftwo
 \fi
}%
\providecommand \natexlab [1]{#1}%
\providecommand \enquote  [1]{``#1''}%
\providecommand \bibnamefont  [1]{#1}%
\providecommand \bibfnamefont [1]{#1}%
\providecommand \citenamefont [1]{#1}%
\providecommand \href@noop [0]{\@secondoftwo}%
\providecommand \href [0]{\begingroup \@sanitize@url \@href}%
\providecommand \@href[1]{\@@startlink{#1}\@@href}%
\providecommand \@@href[1]{\endgroup#1\@@endlink}%
\providecommand \@sanitize@url [0]{\catcode `\\12\catcode `\$12\catcode
  `\&12\catcode `\#12\catcode `\^12\catcode `\_12\catcode `\%12\relax}%
\providecommand \@@startlink[1]{}%
\providecommand \@@endlink[0]{}%
\providecommand \url  [0]{\begingroup\@sanitize@url \@url }%
\providecommand \@url [1]{\endgroup\@href {#1}{\urlprefix }}%
\providecommand \urlprefix  [0]{URL }%
\providecommand \Eprint [0]{\href }%
\providecommand \doibase [0]{http://dx.doi.org/}%
\providecommand \selectlanguage [0]{\@gobble}%
\providecommand \bibinfo  [0]{\@secondoftwo}%
\providecommand \bibfield  [0]{\@secondoftwo}%
\providecommand \translation [1]{[#1]}%
\providecommand \BibitemOpen [0]{}%
\providecommand \bibitemStop [0]{}%
\providecommand \bibitemNoStop [0]{.\EOS\space}%
\providecommand \EOS [0]{\spacefactor3000\relax}%
\providecommand \BibitemShut  [1]{\csname bibitem#1\endcsname}%
\let\auto@bib@innerbib\@empty
\bibitem [{\citenamefont {Mackenzie}\ and\ \citenamefont
  {Maeno}(2003)}]{Mackenzie_RMP_2003}%
  \BibitemOpen
  \bibfield  {author} {\bibinfo {author} {\bibfnamefont {Andrew~Peter}\
  \bibnamefont {Mackenzie}}\ and\ \bibinfo {author} {\bibfnamefont {Yoshiteru}\
  \bibnamefont {Maeno}},\ }\bibfield  {title} {\enquote {\bibinfo {title} {The
  superconductivity of {Sr}$_2${RuO}$_4$ and the physics of spin-triplet
  pairing},}\ }\href {\doibase 10.1103/RevModPhys.75.657} {\bibfield  {journal}
  {\bibinfo  {journal} {Rev. Mod. Phys.}\ }\textbf {\bibinfo {volume} {75}},\
  \bibinfo {pages} {657--712} (\bibinfo {year} {2003})}\BibitemShut {NoStop}%
\bibitem [{\citenamefont {Maeno}\ \emph {et~al.}(2012)\citenamefont {Maeno},
  \citenamefont {Kittaka}, \citenamefont {Nomura}, \citenamefont {Yonezawa},\
  and\ \citenamefont {Ishida}}]{Maeno_JPSJ_2012}%
  \BibitemOpen
  \bibfield  {author} {\bibinfo {author} {\bibfnamefont {Yoshiteru}\
  \bibnamefont {Maeno}}, \bibinfo {author} {\bibfnamefont {Shunichiro}\
  \bibnamefont {Kittaka}}, \bibinfo {author} {\bibfnamefont {Takuji}\
  \bibnamefont {Nomura}}, \bibinfo {author} {\bibfnamefont {Shingo}\
  \bibnamefont {Yonezawa}}, \ and\ \bibinfo {author} {\bibfnamefont {Kenji}\
  \bibnamefont {Ishida}},\ }\bibfield  {title} {\enquote {\bibinfo {title}
  {Evaluation of spin-triplet superconductivity in {Sr}$_2${RuO}$_4$},}\ }\href
  {\doibase 10.1143/JPSJ.81.011009} {\bibfield  {journal} {\bibinfo  {journal}
  {Journal of the Physical Society of Japan}\ }\textbf {\bibinfo {volume}
  {81}},\ \bibinfo {pages} {011009} (\bibinfo {year} {2012})}\BibitemShut
  {NoStop}%
\bibitem [{\citenamefont {Kallin}\ and\ \citenamefont
  {Berlinsky}(2016)}]{Kallin_IOP_2016}%
  \BibitemOpen
  \bibfield  {author} {\bibinfo {author} {\bibfnamefont {Catherine}\
  \bibnamefont {Kallin}}\ and\ \bibinfo {author} {\bibfnamefont {John}\
  \bibnamefont {Berlinsky}},\ }\bibfield  {title} {\enquote {\bibinfo {title}
  {Chiral superconductors},}\ }\href {\doibase 10.1088/0034-4885/79/5/054502}
  {\bibfield  {journal} {\bibinfo  {journal} {Reports on Progress in Physics}\
  }\textbf {\bibinfo {volume} {79}},\ \bibinfo {pages} {054502} (\bibinfo
  {year} {2016})}\BibitemShut {NoStop}%
\bibitem [{\citenamefont {Mackenzie}\ \emph {et~al.}(2017)\citenamefont
  {Mackenzie}, \citenamefont {Scaffidi}, \citenamefont {Hicks},\ and\
  \citenamefont {Maeno}}]{Mackenzie_npj_2017}%
  \BibitemOpen
  \bibfield  {author} {\bibinfo {author} {\bibfnamefont {Andrew~P.}\
  \bibnamefont {Mackenzie}}, \bibinfo {author} {\bibfnamefont {Thomas}\
  \bibnamefont {Scaffidi}}, \bibinfo {author} {\bibfnamefont {Clifford~W.}\
  \bibnamefont {Hicks}}, \ and\ \bibinfo {author} {\bibfnamefont {Yoshiteru}\
  \bibnamefont {Maeno}},\ }\bibfield  {title} {\enquote {\bibinfo {title} {Even
  odder after twenty-three years: the superconducting order parameter puzzle of
  {Sr}$_2${RuO}$_4$},}\ }\href {https://doi.org/10.1038/s41535-017-0045-4}
  {\bibfield  {journal} {\bibinfo  {journal} {npj Quantum Materials}\ }\textbf
  {\bibinfo {volume} {2}} (\bibinfo {year} {2017})}\BibitemShut {NoStop}%
\bibitem [{\citenamefont {Ishida}\ \emph {et~al.}(1998)\citenamefont {Ishida},
  \citenamefont {Mukuda}, \citenamefont {Kitaoka}, \citenamefont {Asayama},
  \citenamefont {Mao}, \citenamefont {Mori},\ and\ \citenamefont
  {Maeno}}]{Ishida_Nature_1998}%
  \BibitemOpen
  \bibfield  {author} {\bibinfo {author} {\bibfnamefont {K.}~\bibnamefont
  {Ishida}}, \bibinfo {author} {\bibfnamefont {H.}~\bibnamefont {Mukuda}},
  \bibinfo {author} {\bibfnamefont {Y.}~\bibnamefont {Kitaoka}}, \bibinfo
  {author} {\bibfnamefont {K.}~\bibnamefont {Asayama}}, \bibinfo {author}
  {\bibfnamefont {Z.~Q.}\ \bibnamefont {Mao}}, \bibinfo {author} {\bibfnamefont
  {Y.}~\bibnamefont {Mori}}, \ and\ \bibinfo {author} {\bibfnamefont
  {Y.}~\bibnamefont {Maeno}},\ }\bibfield  {title} {\enquote {\bibinfo {title}
  {Spin-triplet superconductivity in {Sr}$_2${RuO}$_4$ identified by 17o
  {K}night shift},}\ }\href {\doibase 10.1038/25315} {\bibfield  {journal}
  {\bibinfo  {journal} {Nature}\ }\textbf {\bibinfo {volume} {396}},\ \bibinfo
  {pages} {658--660} (\bibinfo {year} {1998})}\BibitemShut {NoStop}%
\bibitem [{\citenamefont {Luke}\ \emph {et~al.}(1998)\citenamefont {Luke},
  \citenamefont {Fudamoto}, \citenamefont {Kojima}, \citenamefont {Larkin},
  \citenamefont {Merrin}, \citenamefont {Nachumi}, \citenamefont {Uemura},
  \citenamefont {Maeno}, \citenamefont {Mao}, \citenamefont {Mori},
  \citenamefont {Nakamura},\ and\ \citenamefont {Sigrist}}]{Luke_Nature_1998}%
  \BibitemOpen
  \bibfield  {author} {\bibinfo {author} {\bibfnamefont {G.~M.}\ \bibnamefont
  {Luke}}, \bibinfo {author} {\bibfnamefont {Y.}~\bibnamefont {Fudamoto}},
  \bibinfo {author} {\bibfnamefont {K.~M.}\ \bibnamefont {Kojima}}, \bibinfo
  {author} {\bibfnamefont {M.~I.}\ \bibnamefont {Larkin}}, \bibinfo {author}
  {\bibfnamefont {J.}~\bibnamefont {Merrin}}, \bibinfo {author} {\bibfnamefont
  {B.}~\bibnamefont {Nachumi}}, \bibinfo {author} {\bibfnamefont {Y.~J.}\
  \bibnamefont {Uemura}}, \bibinfo {author} {\bibfnamefont {Y.}~\bibnamefont
  {Maeno}}, \bibinfo {author} {\bibfnamefont {Z.~Q.}\ \bibnamefont {Mao}},
  \bibinfo {author} {\bibfnamefont {Y.}~\bibnamefont {Mori}}, \bibinfo {author}
  {\bibfnamefont {H.}~\bibnamefont {Nakamura}}, \ and\ \bibinfo {author}
  {\bibfnamefont {M.}~\bibnamefont {Sigrist}},\ }\bibfield  {title} {\enquote
  {\bibinfo {title} {Time-reversal symmetry-breaking superconductivity in
  {Sr}$_2${RuO}$_4$},}\ }\href {\doibase 10.1038/29038} {\bibfield  {journal}
  {\bibinfo  {journal} {Nature}\ }\textbf {\bibinfo {volume} {394}},\ \bibinfo
  {pages} {558--561} (\bibinfo {year} {1998})}\BibitemShut {NoStop}%
\bibitem [{\citenamefont {Xia}\ \emph {et~al.}(2006)\citenamefont {Xia},
  \citenamefont {Maeno}, \citenamefont {Beyersdorf}, \citenamefont {Fejer},\
  and\ \citenamefont {Kapitulnik}}]{Xia_PRL_2006}%
  \BibitemOpen
  \bibfield  {author} {\bibinfo {author} {\bibfnamefont {Jing}\ \bibnamefont
  {Xia}}, \bibinfo {author} {\bibfnamefont {Yoshiteru}\ \bibnamefont {Maeno}},
  \bibinfo {author} {\bibfnamefont {Peter~T.}\ \bibnamefont {Beyersdorf}},
  \bibinfo {author} {\bibfnamefont {M.~M.}\ \bibnamefont {Fejer}}, \ and\
  \bibinfo {author} {\bibfnamefont {Aharon}\ \bibnamefont {Kapitulnik}},\
  }\bibfield  {title} {\enquote {\bibinfo {title} {High resolution polar kerr
  effect measurements of {Sr}$_2${RuO}$_4$: Evidence for broken time-reversal
  symmetry in the superconducting state},}\ }\href {\doibase
  10.1103/PhysRevLett.97.167002} {\bibfield  {journal} {\bibinfo  {journal}
  {Phys. Rev. Lett.}\ }\textbf {\bibinfo {volume} {97}},\ \bibinfo {pages}
  {167002} (\bibinfo {year} {2006})}\BibitemShut {NoStop}%
\bibitem [{\citenamefont {Rice}\ and\ \citenamefont
  {Sigrist}(1995)}]{Rice_1995}%
  \BibitemOpen
  \bibfield  {author} {\bibinfo {author} {\bibfnamefont {T~M}\ \bibnamefont
  {Rice}}\ and\ \bibinfo {author} {\bibfnamefont {M}~\bibnamefont {Sigrist}},\
  }\bibfield  {title} {\enquote {\bibinfo {title} {{Sr}$_2${RuO}$_4$: an
  electronic analogue of 3{He}?}}\ }\href {\doibase 10.1088/0953-8984/7/47/002}
  {\bibfield  {journal} {\bibinfo  {journal} {Journal of Physics: Condensed
  Matter}\ }\textbf {\bibinfo {volume} {7}},\ \bibinfo {pages} {L643--L648}
  (\bibinfo {year} {1995})}\BibitemShut {NoStop}%
\bibitem [{\citenamefont {Bonalde}\ \emph {et~al.}(2000)\citenamefont
  {Bonalde}, \citenamefont {Yanoff}, \citenamefont {Salamon}, \citenamefont
  {Van~Harlingen}, \citenamefont {Chia}, \citenamefont {Mao},\ and\
  \citenamefont {Maeno}}]{Bonalde_PRL_2000}%
  \BibitemOpen
  \bibfield  {author} {\bibinfo {author} {\bibfnamefont {I.}~\bibnamefont
  {Bonalde}}, \bibinfo {author} {\bibfnamefont {Brian~D.}\ \bibnamefont
  {Yanoff}}, \bibinfo {author} {\bibfnamefont {M.~B.}\ \bibnamefont {Salamon}},
  \bibinfo {author} {\bibfnamefont {D.~J.}\ \bibnamefont {Van~Harlingen}},
  \bibinfo {author} {\bibfnamefont {E.~M.~E.}\ \bibnamefont {Chia}}, \bibinfo
  {author} {\bibfnamefont {Z.~Q.}\ \bibnamefont {Mao}}, \ and\ \bibinfo
  {author} {\bibfnamefont {Y.}~\bibnamefont {Maeno}},\ }\bibfield  {title}
  {\enquote {\bibinfo {title} {Temperature dependence of the penetration depth
  in {Sr}$_2${RuO}$_4$: Evidence for nodes in the gap function},}\ }\href
  {\doibase 10.1103/PhysRevLett.85.4775} {\bibfield  {journal} {\bibinfo
  {journal} {Phys. Rev. Lett.}\ }\textbf {\bibinfo {volume} {85}},\ \bibinfo
  {pages} {4775--4778} (\bibinfo {year} {2000})}\BibitemShut {NoStop}%
\bibitem [{\citenamefont {NishiZaki}\ \emph {et~al.}(2000)\citenamefont
  {NishiZaki}, \citenamefont {Maeno},\ and\ \citenamefont
  {Mao}}]{Mao_JPSJ_2000}%
  \BibitemOpen
  \bibfield  {author} {\bibinfo {author} {\bibfnamefont {Shuji}\ \bibnamefont
  {NishiZaki}}, \bibinfo {author} {\bibfnamefont {Yoshiteru}\ \bibnamefont
  {Maeno}}, \ and\ \bibinfo {author} {\bibfnamefont {Zhiqiang}\ \bibnamefont
  {Mao}},\ }\bibfield  {title} {\enquote {\bibinfo {title} {Changes in the
  superconducting state of {Sr}$_2${RuO}$_4$ under magnetic fields probed by
  specific heat},}\ }\href {\doibase 10.1143/JPSJ.69.572} {\bibfield  {journal}
  {\bibinfo  {journal} {Journal of the Physical Society of Japan}\ }\textbf
  {\bibinfo {volume} {69}},\ \bibinfo {pages} {572--578} (\bibinfo {year}
  {2000})}\BibitemShut {NoStop}%
\bibitem [{\citenamefont {Deguchi}\ \emph {et~al.}(2004)\citenamefont
  {Deguchi}, \citenamefont {Mao}, \citenamefont {Yaguchi},\ and\ \citenamefont
  {Maeno}}]{Maeno_PRL_2004}%
  \BibitemOpen
  \bibfield  {author} {\bibinfo {author} {\bibfnamefont {K.}~\bibnamefont
  {Deguchi}}, \bibinfo {author} {\bibfnamefont {Z.~Q.}\ \bibnamefont {Mao}},
  \bibinfo {author} {\bibfnamefont {H.}~\bibnamefont {Yaguchi}}, \ and\
  \bibinfo {author} {\bibfnamefont {Y.}~\bibnamefont {Maeno}},\ }\bibfield
  {title} {\enquote {\bibinfo {title} {Gap structure of the spin-triplet
  superconductor {Sr}$_2${RuO}$_4$ determined from the field-orientation
  dependence of the specific heat},}\ }\href {\doibase
  10.1103/PhysRevLett.92.047002} {\bibfield  {journal} {\bibinfo  {journal}
  {Phys. Rev. Lett.}\ }\textbf {\bibinfo {volume} {92}},\ \bibinfo {pages}
  {047002} (\bibinfo {year} {2004})}\BibitemShut {NoStop}%
\bibitem [{\citenamefont {Hassinger}\ \emph {et~al.}(2017)\citenamefont
  {Hassinger}, \citenamefont {Bourgeois-Hope}, \citenamefont {Taniguchi},
  \citenamefont {Ren\'e~de Cotret}, \citenamefont {Grissonnanche},
  \citenamefont {Anwar}, \citenamefont {Maeno}, \citenamefont
  {Doiron-Leyraud},\ and\ \citenamefont {Taillefer}}]{Hassinger_PRX_2017}%
  \BibitemOpen
  \bibfield  {author} {\bibinfo {author} {\bibfnamefont {E.}~\bibnamefont
  {Hassinger}}, \bibinfo {author} {\bibfnamefont {P.}~\bibnamefont
  {Bourgeois-Hope}}, \bibinfo {author} {\bibfnamefont {H.}~\bibnamefont
  {Taniguchi}}, \bibinfo {author} {\bibfnamefont {S.}~\bibnamefont {Ren\'e~de
  Cotret}}, \bibinfo {author} {\bibfnamefont {G.}~\bibnamefont
  {Grissonnanche}}, \bibinfo {author} {\bibfnamefont {M.~S.}\ \bibnamefont
  {Anwar}}, \bibinfo {author} {\bibfnamefont {Y.}~\bibnamefont {Maeno}},
  \bibinfo {author} {\bibfnamefont {N.}~\bibnamefont {Doiron-Leyraud}}, \ and\
  \bibinfo {author} {\bibfnamefont {Louis}\ \bibnamefont {Taillefer}},\
  }\bibfield  {title} {\enquote {\bibinfo {title} {Vertical line nodes in the
  superconducting gap structure of {Sr}$_2${RuO}$_4$},}\ }\href {\doibase
  10.1103/PhysRevX.7.011032} {\bibfield  {journal} {\bibinfo  {journal} {Phys.
  Rev. X}\ }\textbf {\bibinfo {volume} {7}},\ \bibinfo {pages} {011032}
  (\bibinfo {year} {2017})}\BibitemShut {NoStop}%
\bibitem [{\citenamefont {Pustogow}\ \emph {et~al.}(2019)\citenamefont
  {Pustogow}, \citenamefont {Luo}, \citenamefont {Chronister}, \citenamefont
  {Su}, \citenamefont {Sokolov}, \citenamefont {Jerzembeck}, \citenamefont
  {Mackenzie}, \citenamefont {Hicks}, \citenamefont {Kikugawa}, \citenamefont
  {Raghu},\ and\ \citenamefont {et~al.}}]{Pustogow_Nature_2019}%
  \BibitemOpen
  \bibfield  {author} {\bibinfo {author} {\bibfnamefont {A.}~\bibnamefont
  {Pustogow}}, \bibinfo {author} {\bibfnamefont {Yongkang}\ \bibnamefont
  {Luo}}, \bibinfo {author} {\bibfnamefont {A.}~\bibnamefont {Chronister}},
  \bibinfo {author} {\bibfnamefont {Y.-S.}\ \bibnamefont {Su}}, \bibinfo
  {author} {\bibfnamefont {D.~A.}\ \bibnamefont {Sokolov}}, \bibinfo {author}
  {\bibfnamefont {F.}~\bibnamefont {Jerzembeck}}, \bibinfo {author}
  {\bibfnamefont {A.~P.}\ \bibnamefont {Mackenzie}}, \bibinfo {author}
  {\bibfnamefont {C.~W.}\ \bibnamefont {Hicks}}, \bibinfo {author}
  {\bibfnamefont {N.}~\bibnamefont {Kikugawa}}, \bibinfo {author}
  {\bibfnamefont {S.}~\bibnamefont {Raghu}}, \ and\ \bibinfo {author}
  {\bibnamefont {et~al.}},\ }\bibfield  {title} {\enquote {\bibinfo {title}
  {Constraints on the superconducting order parameter in {Sr}$_2${RuO}$_4$ from
  oxygen-17 nuclear magnetic resonance},}\ }\href {\doibase
  10.1038/s41586-019-1596-2} {\bibfield  {journal} {\bibinfo  {journal}
  {Nature}\ }\textbf {\bibinfo {volume} {574}},\ \bibinfo {pages} {72--75}
  (\bibinfo {year} {2019})}\BibitemShut {NoStop}%
\bibitem [{\citenamefont {Ishida}\ \emph {et~al.}(2020)\citenamefont {Ishida},
  \citenamefont {Manago}, \citenamefont {Kinjo},\ and\ \citenamefont
  {Maeno}}]{Ishida_correct}%
  \BibitemOpen
  \bibfield  {author} {\bibinfo {author} {\bibfnamefont {Kenji}\ \bibnamefont
  {Ishida}}, \bibinfo {author} {\bibfnamefont {Masahiro}\ \bibnamefont
  {Manago}}, \bibinfo {author} {\bibfnamefont {Katsuki}\ \bibnamefont {Kinjo}},
  \ and\ \bibinfo {author} {\bibfnamefont {Yoshiteru}\ \bibnamefont {Maeno}},\
  }\bibfield  {title} {\enquote {\bibinfo {title} {{Reduction of the $^{17}$O
  {K}night Shift in the Superconducting State and the Heat-up Effect by NMR
  Pulses} on {Sr}$_2${RuO}$_4$},}\ }\href {\doibase 10.7566/JPSJ.89.034712}
  {\bibfield  {journal} {\bibinfo  {journal} {Journal of the Physical Society
  of Japan}\ }\textbf {\bibinfo {volume} {89}},\ \bibinfo {pages} {034712}
  (\bibinfo {year} {2020})}\BibitemShut {NoStop}%
\bibitem [{\citenamefont {Chronister}\ \emph {et~al.}(2021)\citenamefont
  {Chronister}, \citenamefont {Pustogow}, \citenamefont {Kikugawa},
  \citenamefont {Sokolov}, \citenamefont {Jerzembeck}, \citenamefont {Hicks},
  \citenamefont {Mackenzie}, \citenamefont {Bauer},\ and\ \citenamefont
  {Brown}}]{Brown_2020}%
  \BibitemOpen
  \bibfield  {author} {\bibinfo {author} {\bibfnamefont {Aaron}\ \bibnamefont
  {Chronister}}, \bibinfo {author} {\bibfnamefont {Andrej}\ \bibnamefont
  {Pustogow}}, \bibinfo {author} {\bibfnamefont {Naoki}\ \bibnamefont
  {Kikugawa}}, \bibinfo {author} {\bibfnamefont {Dmitry~A.}\ \bibnamefont
  {Sokolov}}, \bibinfo {author} {\bibfnamefont {Fabian}\ \bibnamefont
  {Jerzembeck}}, \bibinfo {author} {\bibfnamefont {Clifford~W.}\ \bibnamefont
  {Hicks}}, \bibinfo {author} {\bibfnamefont {Andrew~P.}\ \bibnamefont
  {Mackenzie}}, \bibinfo {author} {\bibfnamefont {Eric~D.}\ \bibnamefont
  {Bauer}}, \ and\ \bibinfo {author} {\bibfnamefont {Stuart~E.}\ \bibnamefont
  {Brown}},\ }\bibfield  {title} {\enquote {\bibinfo {title} {Evidence for even
  parity unconventional superconductivity in {Sr}$_2${RuO}$_4$},}\ }\href
  {https://www.pnas.org/content/118/25/e2025313118} {\bibfield  {journal}
  {\bibinfo  {journal} {Proceedings of the National Academy of Sciences}\
  }\textbf {\bibinfo {volume} {118}} (\bibinfo {year} {2021})}\BibitemShut
  {NoStop}%
\bibitem [{\citenamefont {Benhabib}\ \emph {et~al.}(2020)\citenamefont
  {Benhabib}, \citenamefont {Lupien}, \citenamefont {Paul}, \citenamefont
  {Berges}, \citenamefont {Dion}, \citenamefont {Nardone}, \citenamefont
  {Zitouni}, \citenamefont {Mao}, \citenamefont {Maeno}, \citenamefont
  {Georges},\ and\ \citenamefont {et~al.}}]{c66_ref}%
  \BibitemOpen
  \bibfield  {author} {\bibinfo {author} {\bibfnamefont {S.}~\bibnamefont
  {Benhabib}}, \bibinfo {author} {\bibfnamefont {C.}~\bibnamefont {Lupien}},
  \bibinfo {author} {\bibfnamefont {I.}~\bibnamefont {Paul}}, \bibinfo {author}
  {\bibfnamefont {L.}~\bibnamefont {Berges}}, \bibinfo {author} {\bibfnamefont
  {M.}~\bibnamefont {Dion}}, \bibinfo {author} {\bibfnamefont {M.}~\bibnamefont
  {Nardone}}, \bibinfo {author} {\bibfnamefont {A.}~\bibnamefont {Zitouni}},
  \bibinfo {author} {\bibfnamefont {Z.~Q.}\ \bibnamefont {Mao}}, \bibinfo
  {author} {\bibfnamefont {Y.}~\bibnamefont {Maeno}}, \bibinfo {author}
  {\bibfnamefont {A.}~\bibnamefont {Georges}}, \ and\ \bibinfo {author}
  {\bibnamefont {et~al.}},\ }\bibfield  {title} {\enquote {\bibinfo {title}
  {Ultrasound evidence for a two-component superconducting order parameter in
  {Sr}$_2${RuO}$_4$},}\ }\href {\doibase 10.1038/s41567-020-1033-3} {\bibfield
  {journal} {\bibinfo  {journal} {Nature Physics}\ }\textbf {\bibinfo {volume}
  {17}},\ \bibinfo {pages} {194--198} (\bibinfo {year} {2020})}\BibitemShut
  {NoStop}%
\bibitem [{\citenamefont {Ghosh}\ \emph {et~al.}(2020)\citenamefont {Ghosh},
  \citenamefont {Shekhter}, \citenamefont {Jerzembeck}, \citenamefont
  {Kikugawa}, \citenamefont {Sokolov}, \citenamefont {Brando}, \citenamefont
  {Mackenzie}, \citenamefont {Hicks},\ and\ \citenamefont
  {Ramshaw}}]{ghosh2020thermodynamic}%
  \BibitemOpen
  \bibfield  {author} {\bibinfo {author} {\bibfnamefont {Sayak}\ \bibnamefont
  {Ghosh}}, \bibinfo {author} {\bibfnamefont {Arkady}\ \bibnamefont
  {Shekhter}}, \bibinfo {author} {\bibfnamefont {F.}~\bibnamefont
  {Jerzembeck}}, \bibinfo {author} {\bibfnamefont {N.}~\bibnamefont
  {Kikugawa}}, \bibinfo {author} {\bibfnamefont {Dmitry~A.}\ \bibnamefont
  {Sokolov}}, \bibinfo {author} {\bibfnamefont {Manuel}\ \bibnamefont
  {Brando}}, \bibinfo {author} {\bibfnamefont {A.~P.}\ \bibnamefont
  {Mackenzie}}, \bibinfo {author} {\bibfnamefont {Clifford~W.}\ \bibnamefont
  {Hicks}}, \ and\ \bibinfo {author} {\bibfnamefont {B.~J.}\ \bibnamefont
  {Ramshaw}},\ }\bibfield  {title} {\enquote {\bibinfo {title} {Thermodynamic
  evidence for a two-component superconducting order parameter in
  {Sr}$_2${RuO}$_4$},}\ }\href {https://doi.org/10.1038/s41567-020-1032-4}
  {\bibfield  {journal} {\bibinfo  {journal} {Nature Physics}\ }\textbf
  {\bibinfo {volume} {17}},\ \bibinfo {pages} {199–204} (\bibinfo {year}
  {2020})}\BibitemShut {NoStop}%
\bibitem [{\citenamefont {Grinenko}\ \emph {et~al.}(2021)\citenamefont
  {Grinenko}, \citenamefont {Ghosh}, \citenamefont {Sarkar}, \citenamefont
  {Orain}, \citenamefont {Nikitin}, \citenamefont {Elender}, \citenamefont
  {Das}, \citenamefont {Guguchia}, \citenamefont {Br{\"u}ckner}, \citenamefont
  {Barber}, \citenamefont {Park}, \citenamefont {Kikugawa}, \citenamefont
  {Sokolov}, \citenamefont {Bobowski}, \citenamefont {Miyoshi}, \citenamefont
  {Maeno}, \citenamefont {Mackenzie}, \citenamefont {Luetkens}, \citenamefont
  {Hicks},\ and\ \citenamefont {Klauss}}]{Grinenko2021}%
  \BibitemOpen
  \bibfield  {author} {\bibinfo {author} {\bibfnamefont {Vadim}\ \bibnamefont
  {Grinenko}}, \bibinfo {author} {\bibfnamefont {Shreenanda}\ \bibnamefont
  {Ghosh}}, \bibinfo {author} {\bibfnamefont {Rajib}\ \bibnamefont {Sarkar}},
  \bibinfo {author} {\bibfnamefont {Jean-Christophe}\ \bibnamefont {Orain}},
  \bibinfo {author} {\bibfnamefont {Artem}\ \bibnamefont {Nikitin}}, \bibinfo
  {author} {\bibfnamefont {Matthias}\ \bibnamefont {Elender}}, \bibinfo
  {author} {\bibfnamefont {Debarchan}\ \bibnamefont {Das}}, \bibinfo {author}
  {\bibfnamefont {Zurab}\ \bibnamefont {Guguchia}}, \bibinfo {author}
  {\bibfnamefont {Felix}\ \bibnamefont {Br{\"u}ckner}}, \bibinfo {author}
  {\bibfnamefont {Mark~E.}\ \bibnamefont {Barber}}, \bibinfo {author}
  {\bibfnamefont {Joonbum}\ \bibnamefont {Park}}, \bibinfo {author}
  {\bibfnamefont {Naoki}\ \bibnamefont {Kikugawa}}, \bibinfo {author}
  {\bibfnamefont {Dmitry~A.}\ \bibnamefont {Sokolov}}, \bibinfo {author}
  {\bibfnamefont {Jake~S.}\ \bibnamefont {Bobowski}}, \bibinfo {author}
  {\bibfnamefont {Takuto}\ \bibnamefont {Miyoshi}}, \bibinfo {author}
  {\bibfnamefont {Yoshiteru}\ \bibnamefont {Maeno}}, \bibinfo {author}
  {\bibfnamefont {Andrew~P.}\ \bibnamefont {Mackenzie}}, \bibinfo {author}
  {\bibfnamefont {Hubertus}\ \bibnamefont {Luetkens}}, \bibinfo {author}
  {\bibfnamefont {Clifford~W.}\ \bibnamefont {Hicks}}, \ and\ \bibinfo {author}
  {\bibfnamefont {Hans-Henning}\ \bibnamefont {Klauss}},\ }\bibfield  {title}
  {\enquote {\bibinfo {title} {Split superconducting and time-reversal
  symmetry-breaking transitions in {Sr}$_2${RuO}$_4$ under stress},}\ }\href
  {\doibase 10.1038/s41567-021-01182-7} {\bibfield  {journal} {\bibinfo
  {journal} {Nature Physics}\ }\textbf {\bibinfo {volume} {17}},\ \bibinfo
  {pages} {748--754} (\bibinfo {year} {2021})}\BibitemShut {NoStop}%
\bibitem [{\citenamefont {R\o{}mer}\ \emph {et~al.}(2019)\citenamefont
  {R\o{}mer}, \citenamefont {Scherer}, \citenamefont {Eremin}, \citenamefont
  {Hirschfeld},\ and\ \citenamefont {Andersen}}]{Astrid_PRL_2019}%
  \BibitemOpen
  \bibfield  {author} {\bibinfo {author} {\bibfnamefont {A.~T.}\ \bibnamefont
  {R\o{}mer}}, \bibinfo {author} {\bibfnamefont {D.~D.}\ \bibnamefont
  {Scherer}}, \bibinfo {author} {\bibfnamefont {I.~M.}\ \bibnamefont {Eremin}},
  \bibinfo {author} {\bibfnamefont {P.~J.}\ \bibnamefont {Hirschfeld}}, \ and\
  \bibinfo {author} {\bibfnamefont {B.~M.}\ \bibnamefont {Andersen}},\
  }\bibfield  {title} {\enquote {\bibinfo {title} {{K}night shift and leading
  superconducting instability from spin fluctuations in {Sr}$_2${RuO}$_4$},}\
  }\href {\doibase 10.1103/PhysRevLett.123.247001} {\bibfield  {journal}
  {\bibinfo  {journal} {Phys. Rev. Lett.}\ }\textbf {\bibinfo {volume} {123}},\
  \bibinfo {pages} {247001} (\bibinfo {year} {2019})}\BibitemShut {NoStop}%
\bibitem [{\citenamefont {R\o{}ising}\ \emph {et~al.}(2019)\citenamefont
  {R\o{}ising}, \citenamefont {Scaffidi}, \citenamefont {Flicker},
  \citenamefont {Lange},\ and\ \citenamefont {Simon}}]{Scaffidi_PRR_2019}%
  \BibitemOpen
  \bibfield  {author} {\bibinfo {author} {\bibfnamefont {Henrik~S.}\
  \bibnamefont {R\o{}ising}}, \bibinfo {author} {\bibfnamefont {Thomas}\
  \bibnamefont {Scaffidi}}, \bibinfo {author} {\bibfnamefont {Felix}\
  \bibnamefont {Flicker}}, \bibinfo {author} {\bibfnamefont {Gunnar~F.}\
  \bibnamefont {Lange}}, \ and\ \bibinfo {author} {\bibfnamefont {Steven~H.}\
  \bibnamefont {Simon}},\ }\bibfield  {title} {\enquote {\bibinfo {title}
  {Superconducting order of {Sr}$_2${RuO}$_4$ from a three-dimensional
  microscopic model},}\ }\href {\doibase 10.1103/PhysRevResearch.1.033108}
  {\bibfield  {journal} {\bibinfo  {journal} {Phys. Rev. Research}\ }\textbf
  {\bibinfo {volume} {1}},\ \bibinfo {pages} {033108} (\bibinfo {year}
  {2019})}\BibitemShut {NoStop}%
\bibitem [{\citenamefont {Gingras}\ \emph {et~al.}(2019)\citenamefont
  {Gingras}, \citenamefont {Nourafkan}, \citenamefont {Tremblay},\ and\
  \citenamefont {C\^ot\'e}}]{Gingras_PRL_2019}%
  \BibitemOpen
  \bibfield  {author} {\bibinfo {author} {\bibfnamefont {O.}~\bibnamefont
  {Gingras}}, \bibinfo {author} {\bibfnamefont {R.}~\bibnamefont {Nourafkan}},
  \bibinfo {author} {\bibfnamefont {A.-M.~S.}\ \bibnamefont {Tremblay}}, \ and\
  \bibinfo {author} {\bibfnamefont {M.}~\bibnamefont {C\^ot\'e}},\ }\bibfield
  {title} {\enquote {\bibinfo {title} {Superconducting symmetries of
  {Sr}$_2${RuO}$_4$ from first-principles electronic structure},}\ }\href
  {\doibase 10.1103/PhysRevLett.123.217005} {\bibfield  {journal} {\bibinfo
  {journal} {Phys. Rev. Lett.}\ }\textbf {\bibinfo {volume} {123}},\ \bibinfo
  {pages} {217005} (\bibinfo {year} {2019})}\BibitemShut {NoStop}%
\bibitem [{\citenamefont {Suh}\ \emph {et~al.}(2020)\citenamefont {Suh},
  \citenamefont {Menke}, \citenamefont {Brydon}, \citenamefont {Timm},
  \citenamefont {Ramires},\ and\ \citenamefont {Agterberg}}]{suh2019}%
  \BibitemOpen
  \bibfield  {author} {\bibinfo {author} {\bibfnamefont {Han~Gyeol}\
  \bibnamefont {Suh}}, \bibinfo {author} {\bibfnamefont {Henri}\ \bibnamefont
  {Menke}}, \bibinfo {author} {\bibfnamefont {P.~M.~R.}\ \bibnamefont
  {Brydon}}, \bibinfo {author} {\bibfnamefont {Carsten}\ \bibnamefont {Timm}},
  \bibinfo {author} {\bibfnamefont {Aline}\ \bibnamefont {Ramires}}, \ and\
  \bibinfo {author} {\bibfnamefont {Daniel~F.}\ \bibnamefont {Agterberg}},\
  }\bibfield  {title} {\enquote {\bibinfo {title} {Stabilizing even-parity
  chiral superconductivity in {Sr}$_2${RuO}$_4$},}\ }\href {\doibase
  10.1103/PhysRevResearch.2.032023} {\bibfield  {journal} {\bibinfo  {journal}
  {Phys. Rev. Research}\ }\textbf {\bibinfo {volume} {2}},\ \bibinfo {pages}
  {032023} (\bibinfo {year} {2020})}\BibitemShut {NoStop}%
\bibitem [{\citenamefont {Kaba}\ and\ \citenamefont
  {S\'en\'echal}(2019)}]{Kaba19}%
  \BibitemOpen
  \bibfield  {author} {\bibinfo {author} {\bibfnamefont {S.-O.}\ \bibnamefont
  {Kaba}}\ and\ \bibinfo {author} {\bibfnamefont {D.}~\bibnamefont
  {S\'en\'echal}},\ }\bibfield  {title} {\enquote {\bibinfo {title}
  {Group-theoretical classification of superconducting states of strontium
  ruthenate},}\ }\href {\doibase 10.1103/PhysRevB.100.214507} {\bibfield
  {journal} {\bibinfo  {journal} {Phys. Rev. B}\ }\textbf {\bibinfo {volume}
  {100}},\ \bibinfo {pages} {214507} (\bibinfo {year} {2019})}\BibitemShut
  {NoStop}%
\bibitem [{\citenamefont {Ramires}\ and\ \citenamefont
  {Sigrist}(2019)}]{Ramires2019}%
  \BibitemOpen
  \bibfield  {author} {\bibinfo {author} {\bibfnamefont {Aline}\ \bibnamefont
  {Ramires}}\ and\ \bibinfo {author} {\bibfnamefont {Manfred}\ \bibnamefont
  {Sigrist}},\ }\bibfield  {title} {\enquote {\bibinfo {title} {Superconducting
  order parameter of {Sr}$_2${RuO}$_4$: A microscopic perspective},}\ }\href
  {\doibase 10.1103/PhysRevB.100.104501} {\bibfield  {journal} {\bibinfo
  {journal} {Phys. Rev. B}\ }\textbf {\bibinfo {volume} {100}},\ \bibinfo
  {pages} {104501} (\bibinfo {year} {2019})}\BibitemShut {NoStop}%
\bibitem [{\citenamefont {Acharya}\ \emph {et~al.}(2019)\citenamefont
  {Acharya}, \citenamefont {Pashov}, \citenamefont {Weber}, \citenamefont
  {Park}, \citenamefont {Sponza},\ and\ \citenamefont
  {Schilfgaarde}}]{Acharya_CommPhys_2019}%
  \BibitemOpen
  \bibfield  {author} {\bibinfo {author} {\bibfnamefont {Swagata}\ \bibnamefont
  {Acharya}}, \bibinfo {author} {\bibfnamefont {Dimitar}\ \bibnamefont
  {Pashov}}, \bibinfo {author} {\bibfnamefont {C{\'e}dric}\ \bibnamefont
  {Weber}}, \bibinfo {author} {\bibfnamefont {Hyowon}\ \bibnamefont {Park}},
  \bibinfo {author} {\bibfnamefont {Lorenzo}\ \bibnamefont {Sponza}}, \ and\
  \bibinfo {author} {\bibfnamefont {Mark~Van}\ \bibnamefont {Schilfgaarde}},\
  }\bibfield  {title} {\enquote {\bibinfo {title} {Evening out the spin and
  charge parity to increase {T}$_c$ in {Sr}$_2${RuO}$_4$},}\ }\href {\doibase
  10.1038/s42005-019-0254-1} {\bibfield  {journal} {\bibinfo  {journal}
  {Communications Physics}\ }\textbf {\bibinfo {volume} {2}},\ \bibinfo {pages}
  {163} (\bibinfo {year} {2019})}\BibitemShut {NoStop}%
\bibitem [{\citenamefont {Wang}\ \emph {et~al.}(2020)\citenamefont {Wang},
  \citenamefont {Wang},\ and\ \citenamefont {Kallin}}]{Kallin_PRB_2020}%
  \BibitemOpen
  \bibfield  {author} {\bibinfo {author} {\bibfnamefont {Zhiqiang}\
  \bibnamefont {Wang}}, \bibinfo {author} {\bibfnamefont {Xin}\ \bibnamefont
  {Wang}}, \ and\ \bibinfo {author} {\bibfnamefont {Catherine}\ \bibnamefont
  {Kallin}},\ }\bibfield  {title} {\enquote {\bibinfo {title} {Spin-orbit
  coupling and spin-triplet pairing symmetry in {Sr}$_2${RuO}$_4$},}\ }\href
  {\doibase 10.1103/PhysRevB.101.064507} {\bibfield  {journal} {\bibinfo
  {journal} {Phys. Rev. B}\ }\textbf {\bibinfo {volume} {101}},\ \bibinfo
  {pages} {064507} (\bibinfo {year} {2020})}\BibitemShut {NoStop}%
\bibitem [{\citenamefont {R\o{}mer}\ \emph {et~al.}(2020)\citenamefont
  {R\o{}mer}, \citenamefont {Kreisel}, \citenamefont {M\"{u}ller},
  \citenamefont {Hirschfeld}, \citenamefont {Eremin},\ and\ \citenamefont
  {Andersen}}]{Romer2020}%
  \BibitemOpen
  \bibfield  {author} {\bibinfo {author} {\bibfnamefont {Astrid~T.}\
  \bibnamefont {R\o{}mer}}, \bibinfo {author} {\bibfnamefont {Andreas}\
  \bibnamefont {Kreisel}}, \bibinfo {author} {\bibfnamefont {Marvin~A.}\
  \bibnamefont {M\"{u}ller}}, \bibinfo {author} {\bibfnamefont {P.~J.}\
  \bibnamefont {Hirschfeld}}, \bibinfo {author} {\bibfnamefont {Ilya~M.}\
  \bibnamefont {Eremin}}, \ and\ \bibinfo {author} {\bibfnamefont {Brian~M.}\
  \bibnamefont {Andersen}},\ }\bibfield  {title} {\enquote {\bibinfo {title}
  {Theory of strain-induced magnetic order and splitting of {T}$_c$ and
  {T}$_{\textnormal{trsb}}$ in {Sr}$_2${RuO}$_4$},}\ }\href
  {http://dx.doi.org/10.1103/PhysRevB.102.054506} {\bibfield  {journal}
  {\bibinfo  {journal} {Phys. Rev. B}\ }\textbf {\bibinfo {volume} {102}}
  (\bibinfo {year} {2020})}\BibitemShut {NoStop}%
\bibitem [{\citenamefont {R\o{}mer}\ and\ \citenamefont
  {Andersen}(2020)}]{romer2020fluctuationdriven}%
  \BibitemOpen
  \bibfield  {author} {\bibinfo {author} {\bibfnamefont {A.~T.}\ \bibnamefont
  {R\o{}mer}}\ and\ \bibinfo {author} {\bibfnamefont {B.~M.}\ \bibnamefont
  {Andersen}},\ }\bibfield  {title} {\enquote {\bibinfo {title}
  {Fluctuation-driven superconductivity in {Sr}$_2${RuO}$_4$ from weak
  repulsive interactions},}\ }\href {\doibase 10.1142/S0217984920400527}
  {\bibfield  {journal} {\bibinfo  {journal} {Modern Physics Letters B}\
  }\textbf {\bibinfo {volume} {34}},\ \bibinfo {pages} {2040052} (\bibinfo
  {year} {2020})}\BibitemShut {NoStop}%
\bibitem [{\citenamefont {R\o{}mer}\ \emph {et~al.}(2021)\citenamefont
  {R\o{}mer}, \citenamefont {Hirschfeld},\ and\ \citenamefont
  {Andersen}}]{Romer2021}%
  \BibitemOpen
  \bibfield  {author} {\bibinfo {author} {\bibfnamefont {Astrid~T.}\
  \bibnamefont {R\o{}mer}}, \bibinfo {author} {\bibfnamefont {P.~J.}\
  \bibnamefont {Hirschfeld}}, \ and\ \bibinfo {author} {\bibfnamefont
  {Brian~M.}\ \bibnamefont {Andersen}},\ }\bibfield  {title} {\enquote
  {\bibinfo {title} {Superconducting state of {Sr}$_2${RuO}$_4$ in the presence
  of longer-range {C}oulomb interactions},}\ }\href {\doibase
  10.1103/PhysRevB.104.064507} {\bibfield  {journal} {\bibinfo  {journal}
  {Phys. Rev. B}\ }\textbf {\bibinfo {volume} {104}},\ \bibinfo {pages}
  {064507} (\bibinfo {year} {2021})}\BibitemShut {NoStop}%
\bibitem [{\citenamefont {Kivelson}\ \emph {et~al.}(2020)\citenamefont
  {Kivelson}, \citenamefont {Yuan}, \citenamefont {Ramshaw},\ and\
  \citenamefont {Thomale}}]{kivelson_npj}%
  \BibitemOpen
  \bibfield  {author} {\bibinfo {author} {\bibfnamefont {Steven~Allan}\
  \bibnamefont {Kivelson}}, \bibinfo {author} {\bibfnamefont {Andrew~Chang}\
  \bibnamefont {Yuan}}, \bibinfo {author} {\bibfnamefont {Brad}\ \bibnamefont
  {Ramshaw}}, \ and\ \bibinfo {author} {\bibfnamefont {Ronny}\ \bibnamefont
  {Thomale}},\ }\bibfield  {title} {\enquote {\bibinfo {title} {A proposal for
  reconciling diverse experiments on the superconducting state in
  {Sr}$_2${RuO}$_4$},}\ }\href {http://dx.doi.org/10.1038/s41535-020-0245-1}
  {\bibfield  {journal} {\bibinfo  {journal} {npj Quantum Materials}\ }\textbf
  {\bibinfo {volume} {5}} (\bibinfo {year} {2020})}\BibitemShut {NoStop}%
\bibitem [{\citenamefont {Clepkens}\ \emph {et~al.}(2021)\citenamefont
  {Clepkens}, \citenamefont {Lindquist},\ and\ \citenamefont
  {Kee}}]{clepkens2021}%
  \BibitemOpen
  \bibfield  {author} {\bibinfo {author} {\bibfnamefont {Jonathan}\
  \bibnamefont {Clepkens}}, \bibinfo {author} {\bibfnamefont {Austin~W.}\
  \bibnamefont {Lindquist}}, \ and\ \bibinfo {author} {\bibfnamefont
  {Hae-Young}\ \bibnamefont {Kee}},\ }\bibfield  {title} {\enquote {\bibinfo
  {title} {Shadowed triplet pairings in {Hund}'s metals with spin-orbit
  coupling},}\ }\href {\doibase 10.1103/PhysRevResearch.3.013001} {\bibfield
  {journal} {\bibinfo  {journal} {Phys. Rev. Research}\ }\textbf {\bibinfo
  {volume} {3}},\ \bibinfo {pages} {013001} (\bibinfo {year}
  {2021})}\BibitemShut {NoStop}%
\bibitem [{\citenamefont {Willa}\ \emph {et~al.}(2021)\citenamefont {Willa},
  \citenamefont {Hecker}, \citenamefont {Fernandes},\ and\ \citenamefont
  {Schmalian}}]{willa2021}%
  \BibitemOpen
  \bibfield  {author} {\bibinfo {author} {\bibfnamefont {Roland}\ \bibnamefont
  {Willa}}, \bibinfo {author} {\bibfnamefont {Matthias}\ \bibnamefont
  {Hecker}}, \bibinfo {author} {\bibfnamefont {Rafael~M.}\ \bibnamefont
  {Fernandes}}, \ and\ \bibinfo {author} {\bibfnamefont {J\"org}\ \bibnamefont
  {Schmalian}},\ }\bibfield  {title} {\enquote {\bibinfo {title} {Inhomogeneous
  time-reversal symmetry breaking in {Sr}$_2${RuO}$_4$},}\ }\href {\doibase
  10.1103/PhysRevB.104.024511} {\bibfield  {journal} {\bibinfo  {journal}
  {Phys. Rev. B}\ }\textbf {\bibinfo {volume} {104}},\ \bibinfo {pages}
  {024511} (\bibinfo {year} {2021})}\BibitemShut {NoStop}%
\bibitem [{\citenamefont {R\o{}mer}\ \emph {et~al.}(2022)\citenamefont
  {R\o{}mer}, \citenamefont {Maier}, \citenamefont {Kreisel}, \citenamefont
  {Hirschfeld},\ and\ \citenamefont {Andersen}}]{Astrid3D2022}%
  \BibitemOpen
  \bibfield  {author} {\bibinfo {author} {\bibfnamefont {Astrid~T.}\
  \bibnamefont {R\o{}mer}}, \bibinfo {author} {\bibfnamefont {T.~A.}\
  \bibnamefont {Maier}}, \bibinfo {author} {\bibfnamefont {Andreas}\
  \bibnamefont {Kreisel}}, \bibinfo {author} {\bibfnamefont {P.~J.}\
  \bibnamefont {Hirschfeld}}, \ and\ \bibinfo {author} {\bibfnamefont
  {Brian~M.}\ \bibnamefont {Andersen}},\ }\bibfield  {title} {\enquote
  {\bibinfo {title} {{Leading superconducting instabilities in
  three-dimensional models for ${\mathrm{Sr}}_{2}{\mathrm{RuO}}_{4}$}},}\
  }\href {\doibase 10.1103/PhysRevResearch.4.033011} {\bibfield  {journal}
  {\bibinfo  {journal} {Phys. Rev. Res.}\ }\textbf {\bibinfo {volume} {4}},\
  \bibinfo {pages} {033011} (\bibinfo {year} {2022})}\BibitemShut {NoStop}%
\bibitem [{\citenamefont {Roig}\ \emph {et~al.}(2022)\citenamefont {Roig},
  \citenamefont {R\o{}mer}, \citenamefont {Kreisel}, \citenamefont
  {Hirschfeld},\ and\ \citenamefont {Andersen}}]{Merce2022}%
  \BibitemOpen
  \bibfield  {author} {\bibinfo {author} {\bibfnamefont {Merc\`e}\ \bibnamefont
  {Roig}}, \bibinfo {author} {\bibfnamefont {Astrid~T.}\ \bibnamefont
  {R\o{}mer}}, \bibinfo {author} {\bibfnamefont {Andreas}\ \bibnamefont
  {Kreisel}}, \bibinfo {author} {\bibfnamefont {P.~J.}\ \bibnamefont
  {Hirschfeld}}, \ and\ \bibinfo {author} {\bibfnamefont {Brian~M.}\
  \bibnamefont {Andersen}},\ }\bibfield  {title} {\enquote {\bibinfo {title}
  {Superconductivity in multiorbital systems with repulsive interactions:
  Hund's pairing versus spin-fluctuation pairing},}\ }\href {\doibase
  10.1103/PhysRevB.106.L100501} {\bibfield  {journal} {\bibinfo  {journal}
  {Phys. Rev. B}\ }\textbf {\bibinfo {volume} {106}},\ \bibinfo {pages}
  {L100501} (\bibinfo {year} {2022})}\BibitemShut {NoStop}%
\bibitem [{\citenamefont {R\o{}ising}\ \emph {et~al.}(2022)\citenamefont
  {R\o{}ising}, \citenamefont {Wagner}, \citenamefont {Roig}, \citenamefont
  {R\o{}mer},\ and\ \citenamefont {Andersen}}]{Henrik2022}%
  \BibitemOpen
  \bibfield  {author} {\bibinfo {author} {\bibfnamefont {Henrik~S.}\
  \bibnamefont {R\o{}ising}}, \bibinfo {author} {\bibfnamefont {Glenn}\
  \bibnamefont {Wagner}}, \bibinfo {author} {\bibfnamefont {Merc\`e}\
  \bibnamefont {Roig}}, \bibinfo {author} {\bibfnamefont {Astrid~T.}\
  \bibnamefont {R\o{}mer}}, \ and\ \bibinfo {author} {\bibfnamefont {Brian~M.}\
  \bibnamefont {Andersen}},\ }\bibfield  {title} {\enquote {\bibinfo {title}
  {Heat capacity double transitions in time-reversal symmetry broken
  superconductors},}\ }\href {\doibase 10.1103/PhysRevB.106.174518} {\bibfield
  {journal} {\bibinfo  {journal} {Phys. Rev. B}\ }\textbf {\bibinfo {volume}
  {106}},\ \bibinfo {pages} {174518} (\bibinfo {year} {2022})}\BibitemShut
  {NoStop}%
\bibitem [{\citenamefont {Kugler}\ \emph {et~al.}(2020)\citenamefont {Kugler},
  \citenamefont {Zingl}, \citenamefont {Strand}, \citenamefont {Lee},
  \citenamefont {von Delft},\ and\ \citenamefont {Georges}}]{Kugler2020}%
  \BibitemOpen
  \bibfield  {author} {\bibinfo {author} {\bibfnamefont {Fabian~B.}\
  \bibnamefont {Kugler}}, \bibinfo {author} {\bibfnamefont {Manuel}\
  \bibnamefont {Zingl}}, \bibinfo {author} {\bibfnamefont {Hugo U.~R.}\
  \bibnamefont {Strand}}, \bibinfo {author} {\bibfnamefont {Seung-Sup~B.}\
  \bibnamefont {Lee}}, \bibinfo {author} {\bibfnamefont {Jan}\ \bibnamefont
  {von Delft}}, \ and\ \bibinfo {author} {\bibfnamefont {Antoine}\ \bibnamefont
  {Georges}},\ }\bibfield  {title} {\enquote {\bibinfo {title} {Strongly
  correlated materials from a numerical renormalization group perspective: How
  the {F}ermi-liquid state of {Sr}$_2${RuO}$_4${} emerges},}\ }\href {\doibase
  10.1103/PhysRevLett.124.016401} {\bibfield  {journal} {\bibinfo  {journal}
  {Phys. Rev. Lett.}\ }\textbf {\bibinfo {volume} {124}},\ \bibinfo {pages}
  {016401} (\bibinfo {year} {2020})}\BibitemShut {NoStop}%
\bibitem [{\citenamefont {Mravlje}\ \emph {et~al.}(2011)\citenamefont
  {Mravlje}, \citenamefont {Aichhorn}, \citenamefont {Miyake}, \citenamefont
  {Haule}, \citenamefont {Kotliar},\ and\ \citenamefont {Georges}}]{Mravlje11}%
  \BibitemOpen
  \bibfield  {author} {\bibinfo {author} {\bibfnamefont {Jernej}\ \bibnamefont
  {Mravlje}}, \bibinfo {author} {\bibfnamefont {Markus}\ \bibnamefont
  {Aichhorn}}, \bibinfo {author} {\bibfnamefont {Takashi}\ \bibnamefont
  {Miyake}}, \bibinfo {author} {\bibfnamefont {Kristjan}\ \bibnamefont
  {Haule}}, \bibinfo {author} {\bibfnamefont {Gabriel}\ \bibnamefont
  {Kotliar}}, \ and\ \bibinfo {author} {\bibfnamefont {Antoine}\ \bibnamefont
  {Georges}},\ }\bibfield  {title} {\enquote {\bibinfo {title}
  {{Coherence-Incoherence Crossover and the Mass-Renormalization Puzzles in
  {Sr}$_2${RuO}$_4$}},}\ }\href {\doibase 10.1103/PhysRevLett.106.096401}
  {\bibfield  {journal} {\bibinfo  {journal} {Phys. Rev. Lett.}\ }\textbf
  {\bibinfo {volume} {106}},\ \bibinfo {pages} {096401} (\bibinfo {year}
  {2011})}\BibitemShut {NoStop}%
\bibitem [{\citenamefont {Haverkort}\ \emph {et~al.}(2008)\citenamefont
  {Haverkort}, \citenamefont {Elfimov}, \citenamefont {Tjeng}, \citenamefont
  {Sawatzky},\ and\ \citenamefont {Damascelli}}]{Damascelli_PRL_2008}%
  \BibitemOpen
  \bibfield  {author} {\bibinfo {author} {\bibfnamefont {M.~W.}\ \bibnamefont
  {Haverkort}}, \bibinfo {author} {\bibfnamefont {I.~S.}\ \bibnamefont
  {Elfimov}}, \bibinfo {author} {\bibfnamefont {L.~H.}\ \bibnamefont {Tjeng}},
  \bibinfo {author} {\bibfnamefont {G.~A.}\ \bibnamefont {Sawatzky}}, \ and\
  \bibinfo {author} {\bibfnamefont {A.}~\bibnamefont {Damascelli}},\ }\bibfield
   {title} {\enquote {\bibinfo {title} {Strong spin-orbit coupling effects on
  the {Fermi} surface of {Sr}$_2${RuO}$_4$ and {Sr}$_2${RhO}$_4$},}\ }\href
  {\doibase 10.1103/PhysRevLett.101.026406} {\bibfield  {journal} {\bibinfo
  {journal} {Phys. Rev. Lett.}\ }\textbf {\bibinfo {volume} {101}},\ \bibinfo
  {pages} {026406} (\bibinfo {year} {2008})}\BibitemShut {NoStop}%
\bibitem [{\citenamefont {Kim}\ \emph {et~al.}(2018)\citenamefont {Kim},
  \citenamefont {Mravlje}, \citenamefont {Ferrero}, \citenamefont {Parcollet},\
  and\ \citenamefont {Georges}}]{Georges_PRL_2018}%
  \BibitemOpen
  \bibfield  {author} {\bibinfo {author} {\bibfnamefont {Minjae}\ \bibnamefont
  {Kim}}, \bibinfo {author} {\bibfnamefont {Jernej}\ \bibnamefont {Mravlje}},
  \bibinfo {author} {\bibfnamefont {Michel}\ \bibnamefont {Ferrero}}, \bibinfo
  {author} {\bibfnamefont {Olivier}\ \bibnamefont {Parcollet}}, \ and\ \bibinfo
  {author} {\bibfnamefont {Antoine}\ \bibnamefont {Georges}},\ }\bibfield
  {title} {\enquote {\bibinfo {title} {Spin-orbit coupling and electronic
  correlations in {Sr}$_2${RuO}$_4$},}\ }\href {\doibase
  10.1103/PhysRevLett.120.126401} {\bibfield  {journal} {\bibinfo  {journal}
  {Phys. Rev. Lett.}\ }\textbf {\bibinfo {volume} {120}},\ \bibinfo {pages}
  {126401} (\bibinfo {year} {2018})}\BibitemShut {NoStop}%
\bibitem [{\citenamefont {Raghu}\ \emph {et~al.}(2010)\citenamefont {Raghu},
  \citenamefont {Kapitulnik},\ and\ \citenamefont {Kivelson}}]{Raghu_PRL_2010}%
  \BibitemOpen
  \bibfield  {author} {\bibinfo {author} {\bibfnamefont {S.}~\bibnamefont
  {Raghu}}, \bibinfo {author} {\bibfnamefont {A.}~\bibnamefont {Kapitulnik}}, \
  and\ \bibinfo {author} {\bibfnamefont {S.~A.}\ \bibnamefont {Kivelson}},\
  }\bibfield  {title} {\enquote {\bibinfo {title} {Hidden quasi-one-dimensional
  superconductivity in {Sr}$_2${RuO}$_4$},}\ }\href {\doibase
  10.1103/PhysRevLett.105.136401} {\bibfield  {journal} {\bibinfo  {journal}
  {Phys. Rev. Lett.}\ }\textbf {\bibinfo {volume} {105}},\ \bibinfo {pages}
  {136401} (\bibinfo {year} {2010})}\BibitemShut {NoStop}%
\bibitem [{\citenamefont {Scaffidi}\ \emph {et~al.}(2014)\citenamefont
  {Scaffidi}, \citenamefont {Romers},\ and\ \citenamefont
  {Simon}}]{Scaffidi_PRB_2014}%
  \BibitemOpen
  \bibfield  {author} {\bibinfo {author} {\bibfnamefont {Thomas}\ \bibnamefont
  {Scaffidi}}, \bibinfo {author} {\bibfnamefont {Jesper~C.}\ \bibnamefont
  {Romers}}, \ and\ \bibinfo {author} {\bibfnamefont {Steven~H.}\ \bibnamefont
  {Simon}},\ }\bibfield  {title} {\enquote {\bibinfo {title} {Pairing symmetry
  and dominant band in {Sr}$_2${RuO}$_4$},}\ }\href {\doibase
  10.1103/PhysRevB.89.220510} {\bibfield  {journal} {\bibinfo  {journal} {Phys.
  Rev. B}\ }\textbf {\bibinfo {volume} {89}},\ \bibinfo {pages} {220510}
  (\bibinfo {year} {2014})}\BibitemShut {NoStop}%
\bibitem [{\citenamefont {Benhabib}\ \emph {et~al.}(2021)\citenamefont
  {Benhabib}, \citenamefont {Lupien}, \citenamefont {Paul}, \citenamefont
  {Berges}, \citenamefont {Dion}, \citenamefont {Nardone}, \citenamefont
  {Zitouni}, \citenamefont {Mao}, \citenamefont {Maeno}, \citenamefont
  {Georges}, \citenamefont {Taillefer},\ and\ \citenamefont
  {Proust}}]{Benhabib2021}%
  \BibitemOpen
  \bibfield  {author} {\bibinfo {author} {\bibfnamefont {S.}~\bibnamefont
  {Benhabib}}, \bibinfo {author} {\bibfnamefont {C.}~\bibnamefont {Lupien}},
  \bibinfo {author} {\bibfnamefont {I.}~\bibnamefont {Paul}}, \bibinfo {author}
  {\bibfnamefont {L.}~\bibnamefont {Berges}}, \bibinfo {author} {\bibfnamefont
  {M.}~\bibnamefont {Dion}}, \bibinfo {author} {\bibfnamefont {M.}~\bibnamefont
  {Nardone}}, \bibinfo {author} {\bibfnamefont {A.}~\bibnamefont {Zitouni}},
  \bibinfo {author} {\bibfnamefont {Z.~Q.}\ \bibnamefont {Mao}}, \bibinfo
  {author} {\bibfnamefont {Y.}~\bibnamefont {Maeno}}, \bibinfo {author}
  {\bibfnamefont {A.}~\bibnamefont {Georges}}, \bibinfo {author} {\bibfnamefont
  {L.}~\bibnamefont {Taillefer}}, \ and\ \bibinfo {author} {\bibfnamefont
  {C.}~\bibnamefont {Proust}},\ }\bibfield  {title} {\enquote {\bibinfo {title}
  {Ultrasound evidence for a two-component superconducting order parameter in
  {Sr}$_2${RuO}$_4$},}\ }\href {\doibase 10.1038/s41567-020-1033-3} {\bibfield
  {journal} {\bibinfo  {journal} {Nature Physics}\ }\textbf {\bibinfo {volume}
  {17}},\ \bibinfo {pages} {194--198} (\bibinfo {year} {2021})}\BibitemShut
  {NoStop}%
\bibitem [{\citenamefont {Firmo}\ \emph
  {et~al.}(2013{\natexlab{a}})\citenamefont {Firmo}, \citenamefont {Lederer},
  \citenamefont {Lupien}, \citenamefont {Mackenzie}, \citenamefont {Davis},\
  and\ \citenamefont {Kivelson}}]{Firmo2013}%
  \BibitemOpen
  \bibfield  {author} {\bibinfo {author} {\bibfnamefont {I.~A.}\ \bibnamefont
  {Firmo}}, \bibinfo {author} {\bibfnamefont {S.}~\bibnamefont {Lederer}},
  \bibinfo {author} {\bibfnamefont {C.}~\bibnamefont {Lupien}}, \bibinfo
  {author} {\bibfnamefont {A.~P.}\ \bibnamefont {Mackenzie}}, \bibinfo {author}
  {\bibfnamefont {J.~C.}\ \bibnamefont {Davis}}, \ and\ \bibinfo {author}
  {\bibfnamefont {S.~A.}\ \bibnamefont {Kivelson}},\ }\bibfield  {title}
  {\enquote {\bibinfo {title} {Evidence from tunneling spectroscopy for a
  quasi-one-dimensional origin of superconductivity in {Sr}$_2${RuO}$_4$},}\
  }\href {\doibase 10.1103/PhysRevB.88.134521} {\bibfield  {journal} {\bibinfo
  {journal} {Phys. Rev. B}\ }\textbf {\bibinfo {volume} {88}},\ \bibinfo
  {pages} {134521} (\bibinfo {year} {2013}{\natexlab{a}})}\BibitemShut
  {NoStop}%
\bibitem [{\citenamefont {Sharma}\ \emph {et~al.}(2020)\citenamefont {Sharma},
  \citenamefont {Edkins}, \citenamefont {Wang}, \citenamefont {Kostin},
  \citenamefont {Sow}, \citenamefont {Maeno}, \citenamefont {Mackenzie},
  \citenamefont {Davis},\ and\ \citenamefont {Madhavan}}]{Madhavan_PNAS_2020}%
  \BibitemOpen
  \bibfield  {author} {\bibinfo {author} {\bibfnamefont {Rahul}\ \bibnamefont
  {Sharma}}, \bibinfo {author} {\bibfnamefont {Stephen~D.}\ \bibnamefont
  {Edkins}}, \bibinfo {author} {\bibfnamefont {Zhenyu}\ \bibnamefont {Wang}},
  \bibinfo {author} {\bibfnamefont {Andrey}\ \bibnamefont {Kostin}}, \bibinfo
  {author} {\bibfnamefont {Chanchal}\ \bibnamefont {Sow}}, \bibinfo {author}
  {\bibfnamefont {Yoshiteru}\ \bibnamefont {Maeno}}, \bibinfo {author}
  {\bibfnamefont {Andrew~P.}\ \bibnamefont {Mackenzie}}, \bibinfo {author}
  {\bibfnamefont {J.~C.~Séamus}\ \bibnamefont {Davis}}, \ and\ \bibinfo
  {author} {\bibfnamefont {Vidya}\ \bibnamefont {Madhavan}},\ }\bibfield
  {title} {\enquote {\bibinfo {title} {Momentum-resolved superconducting energy
  gaps of {Sr}$_2${RuO}$_4$ from quasiparticle interference imaging},}\ }\href
  {\doibase 10.1073/pnas.1916463117} {\bibfield  {journal} {\bibinfo  {journal}
  {Proceedings of the National Academy of Sciences}\ }\textbf {\bibinfo
  {volume} {117}},\ \bibinfo {pages} {5222--5227} (\bibinfo {year}
  {2020})}\BibitemShut {NoStop}%
\bibitem [{\citenamefont {Nunner}\ \emph {et~al.}(2006)\citenamefont {Nunner},
  \citenamefont {Chen}, \citenamefont {Andersen}, \citenamefont {Melikyan},\
  and\ \citenamefont {Hirschfeld}}]{Nunner2006}%
  \BibitemOpen
  \bibfield  {author} {\bibinfo {author} {\bibfnamefont {Tamara~S.}\
  \bibnamefont {Nunner}}, \bibinfo {author} {\bibfnamefont {Wei}\ \bibnamefont
  {Chen}}, \bibinfo {author} {\bibfnamefont {Brian~M.}\ \bibnamefont
  {Andersen}}, \bibinfo {author} {\bibfnamefont {Ashot}\ \bibnamefont
  {Melikyan}}, \ and\ \bibinfo {author} {\bibfnamefont {P.~J.}\ \bibnamefont
  {Hirschfeld}},\ }\bibfield  {title} {\enquote {\bibinfo {title} {Fourier
  transform spectroscopy of $d$-wave quasiparticles in the presence of atomic
  scale pairing disorder},}\ }\href {\doibase 10.1103/PhysRevB.73.104511}
  {\bibfield  {journal} {\bibinfo  {journal} {Phys. Rev. B}\ }\textbf {\bibinfo
  {volume} {73}},\ \bibinfo {pages} {104511} (\bibinfo {year}
  {2006})}\BibitemShut {NoStop}%
\bibitem [{\citenamefont {Hanaguri}\ \emph {et~al.}(2007)\citenamefont
  {Hanaguri}, \citenamefont {Kohsaka}, \citenamefont {Davis}, \citenamefont
  {Lupien}, \citenamefont {Yamada}, \citenamefont {Azuma}, \citenamefont
  {Takano}, \citenamefont {Ohishi}, \citenamefont {Ono},\ and\ \citenamefont
  {Takagi}}]{Hanaguri_NatPhys_2007}%
  \BibitemOpen
  \bibfield  {author} {\bibinfo {author} {\bibfnamefont {T.}~\bibnamefont
  {Hanaguri}}, \bibinfo {author} {\bibfnamefont {Y.}~\bibnamefont {Kohsaka}},
  \bibinfo {author} {\bibfnamefont {J.~C.}\ \bibnamefont {Davis}}, \bibinfo
  {author} {\bibfnamefont {C.}~\bibnamefont {Lupien}}, \bibinfo {author}
  {\bibfnamefont {I.}~\bibnamefont {Yamada}}, \bibinfo {author} {\bibfnamefont
  {M.}~\bibnamefont {Azuma}}, \bibinfo {author} {\bibfnamefont
  {M.}~\bibnamefont {Takano}}, \bibinfo {author} {\bibfnamefont
  {K.}~\bibnamefont {Ohishi}}, \bibinfo {author} {\bibfnamefont
  {M.}~\bibnamefont {Ono}}, \ and\ \bibinfo {author} {\bibfnamefont
  {H.}~\bibnamefont {Takagi}},\ }\bibfield  {title} {\enquote {\bibinfo {title}
  {Quasiparticle interference and superconducting gap in
  {Ca}$_{2-x}${Na}$_x${CuO}$_2${Cl}$_2$},}\ }\href {\doibase 10.1038/nphys753}
  {\bibfield  {journal} {\bibinfo  {journal} {Nature Physics}\ }\textbf
  {\bibinfo {volume} {3}},\ \bibinfo {pages} {865--871} (\bibinfo {year}
  {2007})}\BibitemShut {NoStop}%
\bibitem [{\citenamefont {Allan}\ \emph {et~al.}(2012)\citenamefont {Allan},
  \citenamefont {Rost}, \citenamefont {Mackenzie}, \citenamefont {Xie},
  \citenamefont {Davis}, \citenamefont {Kihou}, \citenamefont {Lee},
  \citenamefont {Iyo}, \citenamefont {Eisaki},\ and\ \citenamefont
  {Chuang}}]{Allan_Science_2012}%
  \BibitemOpen
  \bibfield  {author} {\bibinfo {author} {\bibfnamefont {M.~P.}\ \bibnamefont
  {Allan}}, \bibinfo {author} {\bibfnamefont {A.~W.}\ \bibnamefont {Rost}},
  \bibinfo {author} {\bibfnamefont {A.~P.}\ \bibnamefont {Mackenzie}}, \bibinfo
  {author} {\bibfnamefont {Yang}\ \bibnamefont {Xie}}, \bibinfo {author}
  {\bibfnamefont {J.~C.}\ \bibnamefont {Davis}}, \bibinfo {author}
  {\bibfnamefont {K.}~\bibnamefont {Kihou}}, \bibinfo {author} {\bibfnamefont
  {C.~H.}\ \bibnamefont {Lee}}, \bibinfo {author} {\bibfnamefont
  {A.}~\bibnamefont {Iyo}}, \bibinfo {author} {\bibfnamefont {H.}~\bibnamefont
  {Eisaki}}, \ and\ \bibinfo {author} {\bibfnamefont {T.-M.}\ \bibnamefont
  {Chuang}},\ }\bibfield  {title} {\enquote {\bibinfo {title} {Anisotropic
  energy gaps of iron-based superconductivity from intraband quasiparticle
  interference in {LiFeAs}},}\ }\href {\doibase 10.1126/science.1218726}
  {\bibfield  {journal} {\bibinfo  {journal} {Science}\ }\textbf {\bibinfo
  {volume} {336}},\ \bibinfo {pages} {563--567} (\bibinfo {year}
  {2012})}\BibitemShut {NoStop}%
\bibitem [{\citenamefont {Sprau}\ \emph {et~al.}(2017)\citenamefont {Sprau},
  \citenamefont {Kostin}, \citenamefont {Kreisel}, \citenamefont {B{\"o}hmer},
  \citenamefont {Taufour}, \citenamefont {Canfield}, \citenamefont {Mukherjee},
  \citenamefont {Hirschfeld}, \citenamefont {Andersen},\ and\ \citenamefont
  {Davis}}]{Sprau2017}%
  \BibitemOpen
  \bibfield  {author} {\bibinfo {author} {\bibfnamefont {P.~O.}\ \bibnamefont
  {Sprau}}, \bibinfo {author} {\bibfnamefont {A.}~\bibnamefont {Kostin}},
  \bibinfo {author} {\bibfnamefont {A.}~\bibnamefont {Kreisel}}, \bibinfo
  {author} {\bibfnamefont {A.~E.}\ \bibnamefont {B{\"o}hmer}}, \bibinfo
  {author} {\bibfnamefont {V.}~\bibnamefont {Taufour}}, \bibinfo {author}
  {\bibfnamefont {P.~C.}\ \bibnamefont {Canfield}}, \bibinfo {author}
  {\bibfnamefont {S.}~\bibnamefont {Mukherjee}}, \bibinfo {author}
  {\bibfnamefont {P.~J.}\ \bibnamefont {Hirschfeld}}, \bibinfo {author}
  {\bibfnamefont {B.~M.}\ \bibnamefont {Andersen}}, \ and\ \bibinfo {author}
  {\bibfnamefont {J.~C.~S{\'e}amus}\ \bibnamefont {Davis}},\ }\bibfield
  {title} {\enquote {\bibinfo {title} {Discovery of orbital-selective cooper
  pairing in {FeSe}},}\ }\href {\doibase 10.1126/science.aal1575} {\bibfield
  {journal} {\bibinfo  {journal} {Science}\ }\textbf {\bibinfo {volume}
  {357}},\ \bibinfo {pages} {75--80} (\bibinfo {year} {2017})}\BibitemShut
  {NoStop}%
\bibitem [{\citenamefont {Kashiwaya}\ \emph {et~al.}(2019)\citenamefont
  {Kashiwaya}, \citenamefont {Saitoh}, \citenamefont {Kashiwaya}, \citenamefont
  {Koyanagi}, \citenamefont {Sato}, \citenamefont {Yada}, \citenamefont
  {Tanaka},\ and\ \citenamefont {Maeno}}]{Maeno_PRB_2019}%
  \BibitemOpen
  \bibfield  {author} {\bibinfo {author} {\bibfnamefont {Satoshi}\ \bibnamefont
  {Kashiwaya}}, \bibinfo {author} {\bibfnamefont {Kohta}\ \bibnamefont
  {Saitoh}}, \bibinfo {author} {\bibfnamefont {Hiromi}\ \bibnamefont
  {Kashiwaya}}, \bibinfo {author} {\bibfnamefont {Masao}\ \bibnamefont
  {Koyanagi}}, \bibinfo {author} {\bibfnamefont {Masatoshi}\ \bibnamefont
  {Sato}}, \bibinfo {author} {\bibfnamefont {Keiji}\ \bibnamefont {Yada}},
  \bibinfo {author} {\bibfnamefont {Yukio}\ \bibnamefont {Tanaka}}, \ and\
  \bibinfo {author} {\bibfnamefont {Yoshiteru}\ \bibnamefont {Maeno}},\
  }\bibfield  {title} {\enquote {\bibinfo {title} {Time-reversal invariant
  superconductivity of {Sr}$_2${RuO}$_4$ revealed by josephson effects},}\
  }\href {\doibase 10.1103/PhysRevB.100.094530} {\bibfield  {journal} {\bibinfo
   {journal} {Phys. Rev. B}\ }\textbf {\bibinfo {volume} {100}},\ \bibinfo
  {pages} {094530} (\bibinfo {year} {2019})}\BibitemShut {NoStop}%
\bibitem [{\citenamefont {Kreisel}\ \emph {et~al.}(2015)\citenamefont
  {Kreisel}, \citenamefont {Choubey}, \citenamefont {Berlijn}, \citenamefont
  {Ku}, \citenamefont {Andersen},\ and\ \citenamefont
  {Hirschfeld}}]{Kreisel_PRL_2015}%
  \BibitemOpen
  \bibfield  {author} {\bibinfo {author} {\bibfnamefont {A.}~\bibnamefont
  {Kreisel}}, \bibinfo {author} {\bibfnamefont {Peayush}\ \bibnamefont
  {Choubey}}, \bibinfo {author} {\bibfnamefont {T.}~\bibnamefont {Berlijn}},
  \bibinfo {author} {\bibfnamefont {W.}~\bibnamefont {Ku}}, \bibinfo {author}
  {\bibfnamefont {B.~M.}\ \bibnamefont {Andersen}}, \ and\ \bibinfo {author}
  {\bibfnamefont {P.~J.}\ \bibnamefont {Hirschfeld}},\ }\bibfield  {title}
  {\enquote {\bibinfo {title} {Interpretation of scanning tunneling
  quasiparticle interference and impurity states in cuprates},}\ }\href
  {\doibase 10.1103/PhysRevLett.114.217002} {\bibfield  {journal} {\bibinfo
  {journal} {Phys. Rev. Lett.}\ }\textbf {\bibinfo {volume} {114}},\ \bibinfo
  {pages} {217002} (\bibinfo {year} {2015})}\BibitemShut {NoStop}%
\bibitem [{\citenamefont {Kreisel}\ \emph {et~al.}(2016)\citenamefont
  {Kreisel}, \citenamefont {Nelson}, \citenamefont {Berlijn}, \citenamefont
  {Ku}, \citenamefont {Aluru}, \citenamefont {Chi}, \citenamefont {Zhou},
  \citenamefont {Singh}, \citenamefont {Wahl}, \citenamefont {Liang},
  \citenamefont {Hardy}, \citenamefont {Bonn}, \citenamefont {Hirschfeld},\
  and\ \citenamefont {Andersen}}]{Kreisel_PRB_2016}%
  \BibitemOpen
  \bibfield  {author} {\bibinfo {author} {\bibfnamefont {A.}~\bibnamefont
  {Kreisel}}, \bibinfo {author} {\bibfnamefont {R.}~\bibnamefont {Nelson}},
  \bibinfo {author} {\bibfnamefont {T.}~\bibnamefont {Berlijn}}, \bibinfo
  {author} {\bibfnamefont {W.}~\bibnamefont {Ku}}, \bibinfo {author}
  {\bibfnamefont {Ramakrishna}\ \bibnamefont {Aluru}}, \bibinfo {author}
  {\bibfnamefont {Shun}\ \bibnamefont {Chi}}, \bibinfo {author} {\bibfnamefont
  {Haibiao}\ \bibnamefont {Zhou}}, \bibinfo {author} {\bibfnamefont {Udai~Raj}\
  \bibnamefont {Singh}}, \bibinfo {author} {\bibfnamefont {Peter}\ \bibnamefont
  {Wahl}}, \bibinfo {author} {\bibfnamefont {Ruixing}\ \bibnamefont {Liang}},
  \bibinfo {author} {\bibfnamefont {Walter~N.}\ \bibnamefont {Hardy}}, \bibinfo
  {author} {\bibfnamefont {D.~A.}\ \bibnamefont {Bonn}}, \bibinfo {author}
  {\bibfnamefont {P.~J.}\ \bibnamefont {Hirschfeld}}, \ and\ \bibinfo {author}
  {\bibfnamefont {Brian~M.}\ \bibnamefont {Andersen}},\ }\bibfield  {title}
  {\enquote {\bibinfo {title} {Towards a quantitative description of tunneling
  conductance of superconductors: Application to {LiFeAs}},}\ }\href {\doibase
  10.1103/PhysRevB.94.224518} {\bibfield  {journal} {\bibinfo  {journal} {Phys.
  Rev. B}\ }\textbf {\bibinfo {volume} {94}},\ \bibinfo {pages} {224518}
  (\bibinfo {year} {2016})}\BibitemShut {NoStop}%
\bibitem [{\citenamefont {Zabolotnyy}\ \emph {et~al.}(2013)\citenamefont
  {Zabolotnyy}, \citenamefont {Evtushinsky}, \citenamefont {Kordyuk},
  \citenamefont {Kim}, \citenamefont {Carleschi}, \citenamefont {Doyle},
  \citenamefont {Fittipaldi}, \citenamefont {Cuoco}, \citenamefont
  {Vecchione},\ and\ \citenamefont {Borisenko}}]{Zabolotny2013}%
  \BibitemOpen
  \bibfield  {author} {\bibinfo {author} {\bibfnamefont {V.B.}\ \bibnamefont
  {Zabolotnyy}}, \bibinfo {author} {\bibfnamefont {D.V.}\ \bibnamefont
  {Evtushinsky}}, \bibinfo {author} {\bibfnamefont {A.A.}\ \bibnamefont
  {Kordyuk}}, \bibinfo {author} {\bibfnamefont {T.K.}\ \bibnamefont {Kim}},
  \bibinfo {author} {\bibfnamefont {E.}~\bibnamefont {Carleschi}}, \bibinfo
  {author} {\bibfnamefont {B.P.}\ \bibnamefont {Doyle}}, \bibinfo {author}
  {\bibfnamefont {R.}~\bibnamefont {Fittipaldi}}, \bibinfo {author}
  {\bibfnamefont {M.}~\bibnamefont {Cuoco}}, \bibinfo {author} {\bibfnamefont
  {A.}~\bibnamefont {Vecchione}}, \ and\ \bibinfo {author} {\bibfnamefont
  {S.V.}\ \bibnamefont {Borisenko}},\ }\bibfield  {title} {\enquote {\bibinfo
  {title} {Renormalized band structure of {Sr}$_2${RuO}$_4$: A quasiparticle
  tight-binding approach},}\ }\href {\doibase
  https://doi.org/10.1016/j.elspec.2013.10.003} {\bibfield  {journal} {\bibinfo
   {journal} {Journal of Electron Spectroscopy and Related Phenomena}\ }\textbf
  {\bibinfo {volume} {191}},\ \bibinfo {pages} {48--53} (\bibinfo {year}
  {2013})}\BibitemShut {NoStop}%
\bibitem [{\citenamefont {Cobo}\ \emph {et~al.}(2016)\citenamefont {Cobo},
  \citenamefont {Ahn}, \citenamefont {Eremin},\ and\ \citenamefont
  {Akbari}}]{Cobo_2016}%
  \BibitemOpen
  \bibfield  {author} {\bibinfo {author} {\bibfnamefont {Sergio}\ \bibnamefont
  {Cobo}}, \bibinfo {author} {\bibfnamefont {Felix}\ \bibnamefont {Ahn}},
  \bibinfo {author} {\bibfnamefont {Ilya}\ \bibnamefont {Eremin}}, \ and\
  \bibinfo {author} {\bibfnamefont {Alireza}\ \bibnamefont {Akbari}},\
  }\bibfield  {title} {\enquote {\bibinfo {title} {Anisotropic spin
  fluctuations in {Sr}$_2${RuO}$_4$ : Role of spin-orbit coupling and induced
  strain},}\ }\href {http://dx.doi.org/10.1103/PhysRevB.94.224507} {\bibfield
  {journal} {\bibinfo  {journal} {Phys. Rev. B}\ }\textbf {\bibinfo {volume}
  {94}} (\bibinfo {year} {2016})}\BibitemShut {NoStop}%
\bibitem [{\citenamefont {Hoffman}(2011)}]{Hoffman2011}%
  \BibitemOpen
  \bibfield  {author} {\bibinfo {author} {\bibfnamefont {Jennifer~E}\
  \bibnamefont {Hoffman}},\ }\bibfield  {title} {\enquote {\bibinfo {title}
  {Spectroscopic scanning tunneling microscopy insights into fe-based
  superconductors},}\ }\href {\doibase 10.1088/0034-4885/74/12/124513}
  {\bibfield  {journal} {\bibinfo  {journal} {Reports on Progress in Physics}\
  }\textbf {\bibinfo {volume} {74}},\ \bibinfo {pages} {124513} (\bibinfo
  {year} {2011})}\BibitemShut {NoStop}%
\bibitem [{\citenamefont {Kreisel}\ \emph {et~al.}(2021)\citenamefont
  {Kreisel}, \citenamefont {Marques}, \citenamefont {Rhodes}, \citenamefont
  {Kong}, \citenamefont {Berlijn}, \citenamefont {Fittipaldi}, \citenamefont
  {Granata}, \citenamefont {Vecchione}, \citenamefont {Wahl},\ and\
  \citenamefont {Hirschfeld}}]{kreisel2021unveiling}%
  \BibitemOpen
  \bibfield  {author} {\bibinfo {author} {\bibfnamefont {A.}~\bibnamefont
  {Kreisel}}, \bibinfo {author} {\bibfnamefont {C.~A.}\ \bibnamefont
  {Marques}}, \bibinfo {author} {\bibfnamefont {L.~C.}\ \bibnamefont {Rhodes}},
  \bibinfo {author} {\bibfnamefont {X.}~\bibnamefont {Kong}}, \bibinfo {author}
  {\bibfnamefont {T.}~\bibnamefont {Berlijn}}, \bibinfo {author} {\bibfnamefont
  {R.}~\bibnamefont {Fittipaldi}}, \bibinfo {author} {\bibfnamefont
  {V.}~\bibnamefont {Granata}}, \bibinfo {author} {\bibfnamefont
  {A.}~\bibnamefont {Vecchione}}, \bibinfo {author} {\bibfnamefont
  {P.}~\bibnamefont {Wahl}}, \ and\ \bibinfo {author} {\bibfnamefont {P.~J.}\
  \bibnamefont {Hirschfeld}},\ }\bibfield  {title} {\enquote {\bibinfo {title}
  {{Quasi-particle interference of the van Hove singularity in Sr2RuO4}},}\
  }\href {\doibase 10.1038/s41535-021-00401-x} {\bibfield  {journal} {\bibinfo
  {journal} {npj Quantum Materials}\ }\textbf {\bibinfo {volume} {6}},\
  \bibinfo {pages} {100} (\bibinfo {year} {2021})}\BibitemShut {NoStop}%
\bibitem [{\citenamefont {Firmo}\ \emph
  {et~al.}(2013{\natexlab{b}})\citenamefont {Firmo}, \citenamefont {Lederer},
  \citenamefont {Lupien}, \citenamefont {Mackenzie}, \citenamefont {Davis},\
  and\ \citenamefont {Kivelson}}]{Kivelson_PRB_2013}%
  \BibitemOpen
  \bibfield  {author} {\bibinfo {author} {\bibfnamefont {I.~A.}\ \bibnamefont
  {Firmo}}, \bibinfo {author} {\bibfnamefont {S.}~\bibnamefont {Lederer}},
  \bibinfo {author} {\bibfnamefont {C.}~\bibnamefont {Lupien}}, \bibinfo
  {author} {\bibfnamefont {A.~P.}\ \bibnamefont {Mackenzie}}, \bibinfo {author}
  {\bibfnamefont {J.~C.}\ \bibnamefont {Davis}}, \ and\ \bibinfo {author}
  {\bibfnamefont {S.~A.}\ \bibnamefont {Kivelson}},\ }\bibfield  {title}
  {\enquote {\bibinfo {title} {Evidence from tunneling spectroscopy for a
  quasi-one-dimensional origin of superconductivity in {Sr}$_2${RuO}$_4$},}\
  }\href {\doibase 10.1103/PhysRevB.88.134521} {\bibfield  {journal} {\bibinfo
  {journal} {Phys. Rev. B}\ }\textbf {\bibinfo {volume} {88}},\ \bibinfo
  {pages} {134521} (\bibinfo {year} {2013}{\natexlab{b}})}\BibitemShut
  {NoStop}%
\bibitem [{\citenamefont {Tsuchiura}\ \emph {et~al.}(2001)\citenamefont
  {Tsuchiura}, \citenamefont {Tanaka}, \citenamefont {Ogata},\ and\
  \citenamefont {Kashiwaya}}]{Tsuchiura2001}%
  \BibitemOpen
  \bibfield  {author} {\bibinfo {author} {\bibfnamefont {Hiroki}\ \bibnamefont
  {Tsuchiura}}, \bibinfo {author} {\bibfnamefont {Yukio}\ \bibnamefont
  {Tanaka}}, \bibinfo {author} {\bibfnamefont {Masao}\ \bibnamefont {Ogata}}, \
  and\ \bibinfo {author} {\bibfnamefont {Satoshi}\ \bibnamefont {Kashiwaya}},\
  }\bibfield  {title} {\enquote {\bibinfo {title} {{Local magnetic moments
  around a nonmagnetic impurity in the two-dimensional $t\ensuremath{-}J$
  model}},}\ }\href {\doibase 10.1103/PhysRevB.64.140501} {\bibfield  {journal}
  {\bibinfo  {journal} {Phys. Rev. B}\ }\textbf {\bibinfo {volume} {64}},\
  \bibinfo {pages} {140501} (\bibinfo {year} {2001})}\BibitemShut {NoStop}%
\bibitem [{\citenamefont {Wang}\ and\ \citenamefont {Lee}(2002)}]{ZWang2002}%
  \BibitemOpen
  \bibfield  {author} {\bibinfo {author} {\bibfnamefont {Ziqiang}\ \bibnamefont
  {Wang}}\ and\ \bibinfo {author} {\bibfnamefont {Patrick~A.}\ \bibnamefont
  {Lee}},\ }\bibfield  {title} {\enquote {\bibinfo {title} {{Local Moment
  Formation in the Superconducting State of a Doped Mott Insulator}},}\ }\href
  {\doibase 10.1103/PhysRevLett.89.217002} {\bibfield  {journal} {\bibinfo
  {journal} {Phys. Rev. Lett.}\ }\textbf {\bibinfo {volume} {89}},\ \bibinfo
  {pages} {217002} (\bibinfo {year} {2002})}\BibitemShut {NoStop}%
\bibitem [{\citenamefont {Zhu}\ \emph {et~al.}(2002)\citenamefont {Zhu},
  \citenamefont {Martin},\ and\ \citenamefont {Bishop}}]{Zhu2002}%
  \BibitemOpen
  \bibfield  {author} {\bibinfo {author} {\bibfnamefont {Jian-Xin}\
  \bibnamefont {Zhu}}, \bibinfo {author} {\bibfnamefont {Ivar}\ \bibnamefont
  {Martin}}, \ and\ \bibinfo {author} {\bibfnamefont {A.~R.}\ \bibnamefont
  {Bishop}},\ }\bibfield  {title} {\enquote {\bibinfo {title} {{Spin and Charge
  Order around Vortices and Impurities in High-${T}_{c}$ Superconductors}},}\
  }\href {\doibase 10.1103/PhysRevLett.89.067003} {\bibfield  {journal}
  {\bibinfo  {journal} {Phys. Rev. Lett.}\ }\textbf {\bibinfo {volume} {89}},\
  \bibinfo {pages} {067003} (\bibinfo {year} {2002})}\BibitemShut {NoStop}%
\bibitem [{\citenamefont {Chen}\ and\ \citenamefont {Ting}(2004)}]{Chen2004}%
  \BibitemOpen
  \bibfield  {author} {\bibinfo {author} {\bibfnamefont {Yan}\ \bibnamefont
  {Chen}}\ and\ \bibinfo {author} {\bibfnamefont {C.~S.}\ \bibnamefont
  {Ting}},\ }\bibfield  {title} {\enquote {\bibinfo {title} {{States of Local
  Moment Induced by Nonmagnetic Impurities in Cuprate Superconductors}},}\
  }\href {\doibase 10.1103/PhysRevLett.92.077203} {\bibfield  {journal}
  {\bibinfo  {journal} {Phys. Rev. Lett.}\ }\textbf {\bibinfo {volume} {92}},\
  \bibinfo {pages} {077203} (\bibinfo {year} {2004})}\BibitemShut {NoStop}%
\bibitem [{\citenamefont {Andersen}\ \emph {et~al.}(2007)\citenamefont
  {Andersen}, \citenamefont {Hirschfeld}, \citenamefont {Kampf},\ and\
  \citenamefont {Schmid}}]{Andersen2007}%
  \BibitemOpen
  \bibfield  {author} {\bibinfo {author} {\bibfnamefont {Brian~M.}\
  \bibnamefont {Andersen}}, \bibinfo {author} {\bibfnamefont {P.~J.}\
  \bibnamefont {Hirschfeld}}, \bibinfo {author} {\bibfnamefont {Arno~P.}\
  \bibnamefont {Kampf}}, \ and\ \bibinfo {author} {\bibfnamefont {Markus}\
  \bibnamefont {Schmid}},\ }\bibfield  {title} {\enquote {\bibinfo {title}
  {{Disorder-Induced Static Antiferromagnetism in Cuprate Superconductors}},}\
  }\href {\doibase 10.1103/PhysRevLett.99.147002} {\bibfield  {journal}
  {\bibinfo  {journal} {Phys. Rev. Lett.}\ }\textbf {\bibinfo {volume} {99}},\
  \bibinfo {pages} {147002} (\bibinfo {year} {2007})}\BibitemShut {NoStop}%
\bibitem [{\citenamefont {Harter}\ \emph {et~al.}(2007)\citenamefont {Harter},
  \citenamefont {Andersen}, \citenamefont {Bobroff}, \citenamefont {Gabay},\
  and\ \citenamefont {Hirschfeld}}]{Harter2007}%
  \BibitemOpen
  \bibfield  {author} {\bibinfo {author} {\bibfnamefont {J.~W.}\ \bibnamefont
  {Harter}}, \bibinfo {author} {\bibfnamefont {B.~M.}\ \bibnamefont
  {Andersen}}, \bibinfo {author} {\bibfnamefont {J.}~\bibnamefont {Bobroff}},
  \bibinfo {author} {\bibfnamefont {M.}~\bibnamefont {Gabay}}, \ and\ \bibinfo
  {author} {\bibfnamefont {P.~J.}\ \bibnamefont {Hirschfeld}},\ }\bibfield
  {title} {\enquote {\bibinfo {title} {{Antiferromagnetic correlations and
  impurity broadening of NMR linewidths in cuprate superconductors}},}\ }\href
  {\doibase 10.1103/PhysRevB.75.054520} {\bibfield  {journal} {\bibinfo
  {journal} {Phys. Rev. B}\ }\textbf {\bibinfo {volume} {75}},\ \bibinfo
  {pages} {054520} (\bibinfo {year} {2007})}\BibitemShut {NoStop}%
\bibitem [{\citenamefont {Andersen}\ \emph {et~al.}(2010)\citenamefont
  {Andersen}, \citenamefont {Graser},\ and\ \citenamefont
  {Hirschfeld}}]{Andersen2010}%
  \BibitemOpen
  \bibfield  {author} {\bibinfo {author} {\bibfnamefont {Brian~M.}\
  \bibnamefont {Andersen}}, \bibinfo {author} {\bibfnamefont {Siegfried}\
  \bibnamefont {Graser}}, \ and\ \bibinfo {author} {\bibfnamefont {P.~J.}\
  \bibnamefont {Hirschfeld}},\ }\bibfield  {title} {\enquote {\bibinfo {title}
  {{Disorder-Induced Freezing of Dynamical Spin Fluctuations in Underdoped
  Cuprate Superconductors}},}\ }\href {\doibase 10.1103/PhysRevLett.105.147002}
  {\bibfield  {journal} {\bibinfo  {journal} {Phys. Rev. Lett.}\ }\textbf
  {\bibinfo {volume} {105}},\ \bibinfo {pages} {147002} (\bibinfo {year}
  {2010})}\BibitemShut {NoStop}%
\bibitem [{\citenamefont {Schmid}\ \emph {et~al.}(2010)\citenamefont {Schmid},
  \citenamefont {Andersen}, \citenamefont {Kampf},\ and\ \citenamefont
  {Hirschfeld}}]{Schmid_2010}%
  \BibitemOpen
  \bibfield  {author} {\bibinfo {author} {\bibfnamefont {Markus}\ \bibnamefont
  {Schmid}}, \bibinfo {author} {\bibfnamefont {Brian~M}\ \bibnamefont
  {Andersen}}, \bibinfo {author} {\bibfnamefont {Arno~P}\ \bibnamefont
  {Kampf}}, \ and\ \bibinfo {author} {\bibfnamefont {P~J}\ \bibnamefont
  {Hirschfeld}},\ }\bibfield  {title} {\enquote {\bibinfo {title} {{d-Wave
  superconductivity as a catalyst for antiferromagnetism in underdoped
  cuprates}},}\ }\href {\doibase 10.1088/1367-2630/12/5/053043} {\bibfield
  {journal} {\bibinfo  {journal} {New Journal of Physics}\ }\textbf {\bibinfo
  {volume} {12}},\ \bibinfo {pages} {053043} (\bibinfo {year}
  {2010})}\BibitemShut {NoStop}%
\bibitem [{\citenamefont {Gastiasoro}\ \emph {et~al.}(2013)\citenamefont
  {Gastiasoro}, \citenamefont {Hirschfeld},\ and\ \citenamefont
  {Andersen}}]{Gastiasoro2013}%
  \BibitemOpen
  \bibfield  {author} {\bibinfo {author} {\bibfnamefont {Maria~N.}\
  \bibnamefont {Gastiasoro}}, \bibinfo {author} {\bibfnamefont {P.~J.}\
  \bibnamefont {Hirschfeld}}, \ and\ \bibinfo {author} {\bibfnamefont
  {Brian~M.}\ \bibnamefont {Andersen}},\ }\bibfield  {title} {\enquote
  {\bibinfo {title} {{Impurity states and cooperative magnetic order in
  Fe-based superconductors}},}\ }\href {\doibase 10.1103/PhysRevB.88.220509}
  {\bibfield  {journal} {\bibinfo  {journal} {Phys. Rev. B}\ }\textbf {\bibinfo
  {volume} {88}},\ \bibinfo {pages} {220509} (\bibinfo {year}
  {2013})}\BibitemShut {NoStop}%
\bibitem [{\citenamefont {Zinkl}\ and\ \citenamefont
  {Sigrist}(2021)}]{Sigrist_2021}%
  \BibitemOpen
  \bibfield  {author} {\bibinfo {author} {\bibfnamefont {Bastian}\ \bibnamefont
  {Zinkl}}\ and\ \bibinfo {author} {\bibfnamefont {Manfred}\ \bibnamefont
  {Sigrist}},\ }\bibfield  {title} {\enquote {\bibinfo {title}
  {{Impurity-induced magnetic ordering in
  ${\mathrm{Sr}}_{2}\mathrm{Ru}{\mathrm{O}}_{4}$}},}\ }\href {\doibase
  10.1103/PhysRevResearch.3.023067} {\bibfield  {journal} {\bibinfo  {journal}
  {Phys. Rev. Research}\ }\textbf {\bibinfo {volume} {3}},\ \bibinfo {pages}
  {023067} (\bibinfo {year} {2021})}\BibitemShut {NoStop}%
\bibitem [{\citenamefont {Wang}\ \emph {et~al.}(2017)\citenamefont {Wang},
  \citenamefont {Walkup}, \citenamefont {Derry}, \citenamefont {Scaffidi},
  \citenamefont {Rak}, \citenamefont {Vig}, \citenamefont {Kogar},
  \citenamefont {Zeljkovic}, \citenamefont {Husain}, \citenamefont {Santos},\
  and\ \citenamefont {et~al.}}]{Wang_2017}%
  \BibitemOpen
  \bibfield  {author} {\bibinfo {author} {\bibfnamefont {Zhenyu}\ \bibnamefont
  {Wang}}, \bibinfo {author} {\bibfnamefont {Daniel}\ \bibnamefont {Walkup}},
  \bibinfo {author} {\bibfnamefont {Philip}\ \bibnamefont {Derry}}, \bibinfo
  {author} {\bibfnamefont {Thomas}\ \bibnamefont {Scaffidi}}, \bibinfo {author}
  {\bibfnamefont {Melinda}\ \bibnamefont {Rak}}, \bibinfo {author}
  {\bibfnamefont {Sean}\ \bibnamefont {Vig}}, \bibinfo {author} {\bibfnamefont
  {Anshul}\ \bibnamefont {Kogar}}, \bibinfo {author} {\bibfnamefont {Ilija}\
  \bibnamefont {Zeljkovic}}, \bibinfo {author} {\bibfnamefont {Ali}\
  \bibnamefont {Husain}}, \bibinfo {author} {\bibfnamefont {Luiz~H.}\
  \bibnamefont {Santos}}, \ and\ \bibinfo {author} {\bibnamefont {et~al.}},\
  }\bibfield  {title} {\enquote {\bibinfo {title} {Quasiparticle interference
  and strong electron-mode coupling in the quasi-one-dimensional bands of
  {Sr}$_2${RuO}$_4$},}\ }\href {\doibase 10.1038/nphys4107} {\bibfield
  {journal} {\bibinfo  {journal} {Nature Physics}\ }\textbf {\bibinfo {volume}
  {13}},\ \bibinfo {pages} {799--805} (\bibinfo {year} {2017})}\BibitemShut
  {NoStop}%
\bibitem [{\citenamefont {{McElroy}}\ \emph {et~al.}(2003)\citenamefont
  {{McElroy}}, \citenamefont {{Simmonds}}, \citenamefont {{Hoffman}},
  \citenamefont {{Lee}}, \citenamefont {{Orenstein}}, \citenamefont {{Eisaki}},
  \citenamefont {{Uchida}},\ and\ \citenamefont {{Davis}}}]{McElroy2003}%
  \BibitemOpen
  \bibfield  {author} {\bibinfo {author} {\bibfnamefont {K.}~\bibnamefont
  {{McElroy}}}, \bibinfo {author} {\bibfnamefont {R.~W.}\ \bibnamefont
  {{Simmonds}}}, \bibinfo {author} {\bibfnamefont {J.~E.}\ \bibnamefont
  {{Hoffman}}}, \bibinfo {author} {\bibfnamefont {D.~H.}\ \bibnamefont
  {{Lee}}}, \bibinfo {author} {\bibfnamefont {J.}~\bibnamefont {{Orenstein}}},
  \bibinfo {author} {\bibfnamefont {H.}~\bibnamefont {{Eisaki}}}, \bibinfo
  {author} {\bibfnamefont {S.}~\bibnamefont {{Uchida}}}, \ and\ \bibinfo
  {author} {\bibfnamefont {J.~C.}\ \bibnamefont {{Davis}}},\ }\bibfield
  {title} {\enquote {\bibinfo {title} {{Relating atomic-scale electronic
  phenomena to wave-like quasiparticle states in superconducting
  Bi$_{2}$Sr$_{2}$CaCu$_{2}$O$_{8+{\ensuremath{\delta}}}$}},}\ }\href {\doibase
  10.1038/nature01496} {\bibfield  {journal} {\bibinfo  {journal} {Nature}\
  }\textbf {\bibinfo {volume} {422}},\ \bibinfo {pages} {592--596} (\bibinfo
  {year} {2003})}\BibitemShut {NoStop}%
\bibitem [{\citenamefont {Zhang}\ \emph {et~al.}(2019)\citenamefont {Zhang},
  \citenamefont {Mesaros}, \citenamefont {Fujita}, \citenamefont {Edkins},
  \citenamefont {Hamidian}, \citenamefont {Ch’ng}, \citenamefont {Eisaki},
  \citenamefont {Uchida}, \citenamefont {Davis}, \citenamefont {Khatami},\ and\
  \citenamefont {Kim}}]{Zhang2019_EQM}%
  \BibitemOpen
  \bibfield  {author} {\bibinfo {author} {\bibfnamefont {Yi}~\bibnamefont
  {Zhang}}, \bibinfo {author} {\bibfnamefont {A.}~\bibnamefont {Mesaros}},
  \bibinfo {author} {\bibfnamefont {K.}~\bibnamefont {Fujita}}, \bibinfo
  {author} {\bibfnamefont {S.~D.}\ \bibnamefont {Edkins}}, \bibinfo {author}
  {\bibfnamefont {M.~H.}\ \bibnamefont {Hamidian}}, \bibinfo {author}
  {\bibfnamefont {K.}~\bibnamefont {Ch’ng}}, \bibinfo {author} {\bibfnamefont
  {H.}~\bibnamefont {Eisaki}}, \bibinfo {author} {\bibfnamefont
  {S.}~\bibnamefont {Uchida}}, \bibinfo {author} {\bibfnamefont
  {J.~C.~S\'eamus}\ \bibnamefont {Davis}}, \bibinfo {author} {\bibfnamefont
  {Ehsan}\ \bibnamefont {Khatami}}, \ and\ \bibinfo {author} {\bibfnamefont
  {Eun-Ah}\ \bibnamefont {Kim}},\ }\bibfield  {title} {\enquote {\bibinfo
  {title} {Machine learning in electronic-quantum-matter imaging
  experiments},}\ }\href {\doibase 10.1038/s41586-019-1319-8} {\bibfield
  {journal} {\bibinfo  {journal} {Nature}\ }\textbf {\bibinfo {volume} {570}},\
  \bibinfo {pages} {484--490} (\bibinfo {year} {2019})}\BibitemShut {NoStop}%
\bibitem [{\citenamefont {Kresse}\ and\ \citenamefont
  {Furthm\"uller}(1996)}]{Kresse1996}%
  \BibitemOpen
  \bibfield  {author} {\bibinfo {author} {\bibfnamefont {G.}~\bibnamefont
  {Kresse}}\ and\ \bibinfo {author} {\bibfnamefont {J.}~\bibnamefont
  {Furthm\"uller}},\ }\bibfield  {title} {\enquote {\bibinfo {title} {Efficient
  iterative schemes for ab initio total-energy calculations using a plane-wave
  basis set},}\ }\href {\doibase 10.1103/PhysRevB.54.11169} {\bibfield
  {journal} {\bibinfo  {journal} {Phys. Rev. B}\ }\textbf {\bibinfo {volume}
  {54}},\ \bibinfo {pages} {11169--11186} (\bibinfo {year} {1996})}\BibitemShut
  {NoStop}%
\bibitem [{\citenamefont {Kresse}\ and\ \citenamefont
  {Joubert}(1999)}]{Kresse1999}%
  \BibitemOpen
  \bibfield  {author} {\bibinfo {author} {\bibfnamefont {G.}~\bibnamefont
  {Kresse}}\ and\ \bibinfo {author} {\bibfnamefont {D.}~\bibnamefont
  {Joubert}},\ }\bibfield  {title} {\enquote {\bibinfo {title} {From ultrasoft
  pseudopotentials to the projector augmented-wave method},}\ }\href {\doibase
  10.1103/PhysRevB.59.1758} {\bibfield  {journal} {\bibinfo  {journal} {Phys.
  Rev. B}\ }\textbf {\bibinfo {volume} {59}},\ \bibinfo {pages} {1758--1775}
  (\bibinfo {year} {1999})}\BibitemShut {NoStop}%
\bibitem [{\citenamefont {Perdew}\ \emph {et~al.}(1996)\citenamefont {Perdew},
  \citenamefont {Burke},\ and\ \citenamefont {Ernzerhof}}]{PBE1996}%
  \BibitemOpen
  \bibfield  {author} {\bibinfo {author} {\bibfnamefont {John~P.}\ \bibnamefont
  {Perdew}}, \bibinfo {author} {\bibfnamefont {Kieron}\ \bibnamefont {Burke}},
  \ and\ \bibinfo {author} {\bibfnamefont {Matthias}\ \bibnamefont
  {Ernzerhof}},\ }\bibfield  {title} {\enquote {\bibinfo {title} {Generalized
  gradient approximation made simple},}\ }\href {\doibase
  10.1103/PhysRevLett.77.3865} {\bibfield  {journal} {\bibinfo  {journal}
  {Phys. Rev. Lett.}\ }\textbf {\bibinfo {volume} {77}},\ \bibinfo {pages}
  {3865--3868} (\bibinfo {year} {1996})}\BibitemShut {NoStop}%
\bibitem [{\citenamefont {Mostofi}\ \emph {et~al.}(2008)\citenamefont
  {Mostofi}, \citenamefont {Yates}, \citenamefont {Lee}, \citenamefont {Souza},
  \citenamefont {Vanderbilt},\ and\ \citenamefont {Marzari}}]{Mostofi_2008}%
  \BibitemOpen
  \bibfield  {author} {\bibinfo {author} {\bibfnamefont {Arash~A.}\
  \bibnamefont {Mostofi}}, \bibinfo {author} {\bibfnamefont {Jonathan~R.}\
  \bibnamefont {Yates}}, \bibinfo {author} {\bibfnamefont {Young-Su}\
  \bibnamefont {Lee}}, \bibinfo {author} {\bibfnamefont {Ivo}\ \bibnamefont
  {Souza}}, \bibinfo {author} {\bibfnamefont {David}\ \bibnamefont
  {Vanderbilt}}, \ and\ \bibinfo {author} {\bibfnamefont {Nicola}\ \bibnamefont
  {Marzari}},\ }\bibfield  {title} {\enquote {\bibinfo {title} {{Wannier}90: A
  tool for obtaining maximally-localised {Wannier} functions},}\ }\href
  {\doibase 10.1016/j.cpc.2007.11.016} {\bibfield  {journal} {\bibinfo
  {journal} {Computer Physics Communications}\ }\textbf {\bibinfo {volume}
  {178}},\ \bibinfo {pages} {685--699} (\bibinfo {year} {2008})}\BibitemShut
  {NoStop}%
\end{thebibliography}%

\end{document}